\documentstyle[preprint,12pt,aps]{revtex} 
\title{Dynamical properties of small polarons}
\author{E.V.L. de Mello$^\dagger$~and~J.~Ranninger} 
\address{Centre de
Recherches sur les Tr\`es Basses Temp\'eratures, Laboratoire
Associ\'e \'a l'Universit\'e Joseph Fourier, Centre National de la
Recherche Scientifique,\\ BP 166, 38042, Grenoble C\'edex 9, France}
\date{\today} 
\begin{document} 
\maketitle 
\draft 
\begin{abstract}

On the basis of the two-site polaron problem, which we solve by 
exact diagonalization, we analyse the spectral properties of 
polaronic systems in view of discerning localized from itinerant 
polarons and bound polaron pairs from an ensemble of single polarons. 
The corresponding experimental techniques for that concern 
photoemission and inverse photoemission spectroscopy. The evolution 
of the density of states as a function of concentration of charge 
carriers and strength of the electron-phonon interaction clearly 
shows the opening up of a gap between single polaronic and 
bi-polaronic states, in analogy to the Hubbard problem for 
strongly correlated electron systems.
In studying the details of the intricately linked dynamics of 
the charge carriers and of the molecular deformations which 
surround them, we find that in general
the dynamical delocalization of the charge carriers helps to 
strengthen the phase coherence for itinerant polaronic states,  
except for the crossover regime between  adiabatic and 
anti-adiabatic small polarons. The crossover between these  
two regimes is triggered by two characteristic time scales: 
the renormalized electron hopping rate and the renormalized 
vibrational frequency becoming equal. This crossover regime 
is then characterized by temporarily alternating self- 
localization and delocalization of the charge carriers which 
is accompanied by phase slips in the charge and molecular 
deformation oscillations and ultimately leads to a dephasing 
between these two dynamical components of the polaron problem.
We visualize these features by a study of the temporal evolution 
of the charge redistribution and the change in molecular 
deformations. The spectral and dynamical properties of polarons 
discussed here are beyond the applicability of the standard 
Lang Firsov approach to the polaron problem.


\end{abstract} 

\newpage 
\section{INTRODUCTION}
 
Renewed interest in the physics of small polarons in the last few 
years has been largely stimulated by the discovery of new materials 
with exceptional properties such as the High $T_c$ cuprates, the 
nickelates and the manganites showing giant magneto resistance.
It is believed that small polarons play an essential role in these 
materials. Recently developed experimental techniques such as femto- 
second optical spectroscopy, EXAFS and pulsed neutron scattering, 
ionchanneling experiments and high resolution
angle resolved photoemission experiments have been employed to 
study the properties of polaronic systems. Inspite of considerable 
theoretical efforts in the last few years, the underlying physics 
of the polaron problem has remained largely unresolved. 

As concerns the problem of a single polaron in an empty lattice 
with local coupling of the charge to a set of individual 
non-interacting local lattice deformations (the Holstein model) 
it is known that self-trapping of a charge carrier occurs when  

i) the gain in localization energy $\varepsilon_p$ outweighs the 
gain in kinetic energy $D/2$, i.e., $g \equiv 2\varepsilon_p /D \geq 1$, 
$D$ denoting the bare electron bandwidth and 

ii) the relative deformation of the lattice which surrounds the 
charge carrier essentially remains confined to the immediate 
vicinity of the charge carrier i.e., 
$2\alpha^2 \equiv \varepsilon_p / \omega_0 \geq 1$, $\omega_0$ 
denoting the bare local phonon frequency.

A crossover between essentially delocalized quasi-free electrons 
and self-trapped electrons is known from Quantum Monte Carlo 
simulations \cite{Raedt-83} to occur in a regime of parameters characterized by
the two conditions i) and ii) for small polaron formation, i.e. for 
$g \sim 1$ and $\alpha \sim 1$. This crossover regime in $\alpha$ narrows substantially upon going from the anti-adiabatic 
($\gamma \equiv t/\omega_0  = 4\alpha^2/(zg) \ll 1$) to the adiabatic 
($\gamma \gg 1$) limit with z denoting the coordination number.

The standard theory of small polarons is based on the so called 
Lang Firsov (LF) transformation usually followed by approximations 
treating the fully localized polaron state as the starting point of 
the theory and then introducing the hopping of the electrons by 
perturbative means \cite {LF-62}. There is a widespread (however erroneous) belief that such a scheme is valid in the extreme strong coupling
anti-adiabatic limit i.e., $\gamma \ll 1$ and $\alpha \gg 1$.
It is not surprising that such 
theoretical treatment is even less able to describe polarons 
close to the crossover regime, which represents the  physically 
interesting and realistic situations.

As concerns the problem of the many polaron system the principal 
difficulty is to take into account the overlap of the lattice 
deformations surrounding the charge carriers when the density 
of polarons becomes important. It is expected that, due to 
destructive interference of such local deformations, the whole 
concept of polarons breaks down, the lattice changes its 
structure and as a result the effective electron lattice 
coupling becomes strongly reduced.

It is the purpose of this present work to obtain some insight 
into the physics of small polarons for realistic values of the 
relevant parameters $\gamma$ and $\alpha$ and in particular 
close to the crossover regime. Since all known materials 
containing small polarons consist of highly polarizable 
small clusters ($Ti^{3+}-Ti^{3+}$ pairs in 
$Ti_4O_7$, $\; V^{4+}-V^{4+}$ pairs in $Na_xV_2O_5$), 
sometimes embedded in a metallic background (such as the $O(4)^{2-}-Cu(1)^{+}-O(4)^{2-}$ dumbbells in cuprate High 
$T_c$ superconductors containing chains separating the metalic 
$CuO_2$ layers), it is not only instructive from a pedagogical 
point of view to study small polaron features on the basis of 
such small units, but such studies may also apply to true 
physically relevant situations. On the basis of a two-site 
polaron system (a system consisting of two adjacent molecules 
between which the electrons can hop) we shall illustrate the 
highly non-linear physics going on in such a problem. For 
$\alpha \gg 1$ and $\gamma \ll 1$ (the so called strong coupling 
anti-adiabatic regime or extreme polaron limit) it will be shown 
that the dynamics of the fluctuations of the lattice deformations 
is driven by the  charge fluctuations, while for 
$\alpha \ll 1$ and $\gamma \gg 1$ (the so called weak coupling 
adiabatic regime) the inverse happens. The crossover regime is 
characterized by alternating localization and delocalization of 
the charge carriers as a function of time. In general the 
amplitude and the frequency of the intrinsic lattice vibrations 
as well as of the intrinsic electron hopping rate are strongly 
renormalized. Such effects may turn out to be vital as concerns a 
mechanism for the damping of polaronic charge carriers; a feature 
which is inaccessible within the usual LF approach.

The paper is organized in the following way:
In section 2 we shall present the model and its basic physics as 
well as our method of an exact diagonalization study. In section 3 
we discuss the dependence of the kinetic energy of the electron  
on the adiabaticity parameter $\gamma$ and the coupling strength 
$\alpha$ and examine the limitations of the  LF approach. 
Section 4 will be devoted to the study of the one particle 
spectral function
in view of distinguishing localized from itinerant polarons 
and to discern tightly bound polaron pairs (bipolarons) from 
simple polarons by such methods as photoemission spectroscopy.
In section 5 we will study the time evolution of the charge 
redistribution 
and of the lattice deformation and discuss the renormalization 
of the intrinsic lattice vibrational frequency and of the electron 
hopping integral in the course of the charge transfer process.

\section{THE MODEL}
The smallest system on which polaronic features can be studied as 
far as self-trapping, localization-delocalization crossover, 
frequency renormalization and dephasing of the correlated 
charge-deformation dynamics is concerned, is the so called 
two-site polaron system. It consists of two elastically 
uncoupled adjacent diatomic molecules with atoms having mass 
$M$ and an intrinsic bare frequency of intra-molecular vibrations 
$\omega_0$. Electrons can hop between those two molecules having a 
bare hopping rate $t$. In the Holstein model for such a system, 
the strength of the coupling constant of the density of charge 
carriers to the intra-molecular deformations is denoted by 
$\lambda$. The model Hamiltonian for such a system is then 
given by
\begin{eqnarray}
H &=& t\sum_{\sigma}(n_{1\sigma}+n_{2\sigma}) - 
t\sum_{\sigma}(c^{\dag}_{1\sigma}c^{\phantom{\dag}}_{2\sigma}+
c^{\dag}_{2\sigma}c^{\phantom{\dag}}_{1\sigma})
-\lambda\sum_{\sigma}(n_{1\sigma}u_1 +n_{2\sigma}u_2) \nonumber \\ 
& &+\frac{M}{2}(\dot{u}^2_1+\dot{u}^2_2)+\frac{M}{2}
\omega^2_0(u^2_1+u^2_2)
+ U(n_{1\uparrow}n_{1\downarrow}+n_{2\uparrow}
n_{2\downarrow}) \label{H}
\end{eqnarray}
where $n_{i\sigma} = c^{\dag}_{i\sigma}c^{\phantom{\dag}}_{i\sigma}$ 
denotes the density of charge carriers at molecular site i and having 
spin $\sigma$. The intra-molecular deformations are denoted by $u_1$ 
and $u_2$ respectively.
We assume a simple form of repulsive intra-molecular Coulomb forces 
of strength U. The above Hamiltonian Eq.(\ref{H}) has been extensively 
studied by a number of 
autors\cite{Sho-73,Pre-79,Web-85,Pas-88,Pro-89,Mar-90,Ran-92,Ale-94} 
who, by means of exact diagonalization, 
have examined various aspects of the polaron problem. Diagonalization 
of the above Hamiltonian can be  rendered more 
efficient\cite{Ran-92} when decomposing it into a term 
containing the symmetric in-phase lattice vibrations characterized 
by a wavevector $q = 0$ and antisymmetric out-of-phase vibrations 
characterized by a wavevector $q = \pi$. The Hamiltonian then can 
be separated into two independent contributions $H = H_X + H_Y$, 
given by:
\begin{eqnarray}\label{H_X}
H_X  &=&  t\sum_{\sigma}(n_{1\sigma}+n_{2\sigma}) - 
t\sum_{\sigma}(c^{\dag}_{1\sigma}c^{\phantom{\dag}}_{2\sigma}+
c^{\dag}_{2\sigma}c^{\phantom{\dag}}_{1\sigma}) 
-\lambda/ \sqrt{2}\sum_{\sigma}(n_{1\sigma}-n_{2\sigma})X \nonumber \\ 
& &+\frac{M}{2}\dot{X}^2 +\frac{M}{2}\omega^2_0X^2 + U(n_{1\uparrow}n_{1\downarrow}+n_{2\uparrow}n_{2\downarrow})
\end{eqnarray}
\begin{equation}\label{H_Y}
H_Y  =  \frac{M}{2}\left[ \dot{Y}-\frac{\lambda \dot{n}}{\sqrt2 M 
\omega^2_0}\right]^2 + 
\frac{M}{2}\omega^2_0\left[ Y-\frac{\lambda n}{\sqrt2 M 
\omega^2_0} \right]^2 - \frac{\lambda^2 n^2}{4 M \omega^2_0} 
\end{equation}
where 
\begin{eqnarray}
X & = & \frac{u_1-u_2}{\sqrt{2}} \; = \; 
\frac{a^{\dag}+a}{\sqrt{2 M \omega_0 / \hbar}} \label{X} \\
Y & = & \frac{u_1+u_2}{\sqrt{2}} \; = \; 
\frac{b^{\dag}+b}{\sqrt{2 M \omega_0 / \hbar}}  
+ \frac{\lambda n}{\sqrt{2} M \omega_{0}^{2}}  \label{Y}
\end{eqnarray}
and $a^{(\dag)}$ and $b^{(\dag)}$ denote the annihilation (creation) 
operators of the quantized lattice fluctuations with momentum $q=\pi$ 
and $q=0$ respectively. Since in $H_Y$  the phonons couple only to the 
total charge $n = \sum_{i\sigma} n_{i\sigma}$ of the system, $H_Y$ can 
be diagonalized exactly since it represents simply a shifted oscillator,  corresponding to the two diatomic molecules of our system having their equilibrium positions shifted by equal amounts $u^0 = (\lambda n)/2M \omega^2_0)$.
$H_X$ on the contrary contains the full dynamics of the system and has 
to be diagonalized numerically in terms of a judiciously chosen set of 
basis states (see ref.\cite{Ran-92}) such as
\begin{eqnarray}
|l_X,l_Y \rangle &=& \sum_{N_X,N_Y}^\infty\frac{1}{\sqrt2}
c^{\dagger}_{1,\sigma}(\alpha^{+}_{l_X,N_X} + \alpha^{-}_{l_X,N_X})|N_X\rangle
\beta_{l_Y,N_Y}|N_Y\rangle + \nonumber \\         
& &\sum_{N_X,N_Y}^\infty\frac{1}{\sqrt2}\frac{1}{\sqrt2}c^{\dagger}_{2,\sigma}
(\alpha^{+}_{l_X,N_X} - \alpha^{-}_{l_X,N_X})|N_X\rangle
\beta_{l_Y,N_Y}|N_Y\rangle \label{states}
\end{eqnarray}
for the one-electron two-site problem. $|N_{X,Y}\rangle$ denote 
the eigenstates of the oscillator part of $H_{X,Y}$ for 
$\lambda=0$. The coefficients $\beta_{l_Y,N_Y}$ are known analytically 
from the expansion of a shifted oscilator states in terms of the 
excited states of the unshifted one. For $l_Y=0$ we have $\beta_{0,N_Y}=exp-{(\alpha^2/2)} \alpha^{N_Y}/\sqrt{N_Y!}$.
The coefficients $\alpha^{+,-}_{l_X,l_Y}$ are determined by diagonalyzing numerically
$H_X$ with in a truncated Hilbertspace of  states with up to 100 
phonon states, i.e., $0 \leq N_X \leq 100$.
The procedure of separating off in the 
Hamiltonian a part which can be diagonalized exactly represents a 
substantial reduction in the numerical  work of studying polarons not 
only for our two-site model but in general\cite{Robin-96}. 

Applying the standard LF approach to this problem amounts to using a representation in which the molecules exist in a set of oscillator 
states: those not containing any electron and being described by the 
full set of oscillator states labeled by 
$|\Phi(X)\rangle_m$ and $|\Phi(Y)\rangle_n$, those containing an 
electron on one of the two sites and being described by shifted 
oscillator states 
$|\Phi(X \pm X_0)\rangle_m$ and $|\Phi(Y-Y_0)\rangle_n$, those 
containing two electrons on one of the two sites and being described 
by $|\Phi(X \pm 2X_0)\rangle_m$ and $|\Phi(Y-2Y_0)\rangle_n$ and 
finally those containing one electron on each site and
being described by $|\Phi(X)\rangle_m$ and $|\Phi(Y-2Y_0)\rangle_n$. 
The shift in equilibrium positions for the two oscillators are given 
by $X_0 = Y_0 =
\lambda n/\sqrt 2 M\omega^2_0$ and the shifted oscillator states are 
defined by
\begin{eqnarray}\label{Osc}
|\Phi(X-X_0)\rangle_m &= & \frac{(a^{\dag}-\alpha)^m}{\sqrt{m!}}exp(\alpha[a^{\dag}-a])|\Phi(X)
\rangle_0
\nonumber \\ 
&=& \frac{(a^{\dag}-\alpha)^m}{\sqrt{m!}}
\sum_l \frac{\alpha^l}{\sqrt {l!}} exp(-\alpha^2/2) |\Phi(X)\rangle_l
\end{eqnarray}
In such a representation the Hamiltonian $H_X$ becomes
\begin{eqnarray}\label{H_X1}
H_X  &=&  t\sum_{\sigma}(n_{1\sigma}+n_{2\sigma}) - 
t\sum_{\sigma}(c^{\dag}_{1\sigma}exp(\alpha[a^{\dag}-a])
c^{\phantom{\dag}}_{2\sigma}+H.c.) \nonumber \\
& &- \varepsilon_pn \; - \; (2\varepsilon_p-U) \sum_{i}n_{i\uparrow}n_{i\downarrow} + \hbar \omega_0 (a^{\dag}a+
\frac{1}{2})
\end{eqnarray}
in which the intricate dynamics of the polaron problem is contained 
in the modified hopping term (the second term in Eq.(\ref {H_X1})) 
showing a concomitant transfer of charge and deformation.
The standard LF approach to this problem is then to consider, to within 
a first approximation, that the deformation of the molecules follows instantaneously the motion of the electrons, that is to say without any 
emission or absorbtion of phonons during the process which transfers an 
electron from one site to the other. For our two-site polaron model this 
implies states of the form
\begin {equation}\label{LFstates}
|LF\rangle^{\pm}_{mn}=\frac{1}{\sqrt 2} \left (c^{\dag}_{1 \sigma}|\Phi(X-X_0)\rangle_m \pm
c^{\dag}_{2 \sigma}|\Phi(X+X_0)\rangle_m \right )|\Phi(Y-Y_0)\rangle_n 
\end{equation}
which amounts to disentangle the correlated hopping term in the form

\begin{eqnarray}\label{LF}
c^{\dag}_{1\sigma}exp(\alpha[a^{\dag}-a])
c^{\phantom{\dag}}_{2\sigma} \Rightarrow \nonumber \\    c^{\dag}_{1\sigma}c^{\phantom{\dag}}_{2\sigma} \langle 
exp(\alpha[a^{\dag}-a])
\rangle_{H_{ph}} = t^*_{LF}c^{\dag}_{1\sigma}
c^{\phantom{\dag}}_{2\sigma}
\end{eqnarray}
This way one obtains an effective Hamiltonian to start with and 
subsequently treats the remainder of the full Hamiltonian in a 
perturbative way - corresponding to an expansion in terms of $1/\lambda$\cite{Firsov-75}. The average in Eq.(\ref{LF}) is taken 
over the free phonon Hamiltonian $H_{ph}$, given by the last term 
in Eq.(\ref{H_X1}) and which leads to 
$t^{*}_{LF} = t exp(-2\alpha^2)$. For a many polaron problem, 
polaronic states may become unstable with respect to bi-polaron 
formation provided the Coulomb repulsion $U$ is overcome. This 
then leads to states of the form $c^{\dag}_{i\uparrow}
c^{\dag}_{i\downarrow}
|\Phi(X \pm 2X_0)\rangle_m |\Phi(Y \pm 2Y_0)\rangle_n $. 
Upon eliminating single polaron states 
in the Hamiltonian, Eq(\ref H) one derives \cite{Alexandrov-81} an 
effective Hamiltonian describing bi-polaronic hopping with a 
hopping integral $t^{**}_{LF}=(t^2/(2\varepsilon_p)exp(-4\alpha^2)$ 
for $U=0$.

\section{Intersite charge transfer and multiple hopping processes}

It has remained a question of dispute as how good the LF approach 
is and whether in the extreme polaronic limit the LF approximation,  
i.e. the substitution described by Eq.(\ref{LF}), becomes exact. 
If this is not the case, then a perturbation theory in terms of 
$1/\lambda$ can not be applied to the polaron problem. It has for a long 
time 
been taken as granted that the problem of a single polaron in an 
infinite lattice of finite dimension is described by polaron band 
states having a  $k$ dependent dispersion identical to that of the 
bare electron but with a reduced bandwidth 
$2zt^{*}$, where z denotes the coordination number. There are now 
indications from exact diagonalization studies of finite clusters\cite{Wellein-96} that this is not correct and that the 
$k$ dependence of the dispersion differs significantely from that 
of the bare electrons. From exact diagonalisation 
studies\cite{Ran-92,Ale-94} of the two-site polaron problem 
we know that over a large regime of parameters $\alpha$ and $\gamma$ 
the LF approximation gives reliable results as far as the energies 
of the low lying excitations are concerned. Nevertheless there can 
be serious discrepancies between the LF approach and the exact 
results when considering  the eigenstates of the polaron problem. 
Let us illustrate theses discrepancies in such physical quantities as 

i) the kinetic energy of the electrons, 

ii) the occupation number of electrons as a function of wave vector,

iii) the wave vector dependence of the spectral function measurable 
by angle resolved photoemission experiments,

iv) the importance of renormalization of the phonon frequencies and 
of the electron hopping integral

v) the retardation between the dynamics of the charge carriers and of 
the lattice deformation which accompanies them.

Let us start by considering the static correlation function 
$t_{eff}/2t=- E_{kin}/2t \equiv 
\langle c^{\dag}_{1\sigma}c^{\phantom{\dag}}_{2\sigma}\rangle$
which describes the kinetic energy in units of $2t$ and which in the 
extreme polaronic LF limit (Eq.(\ref{LF})) becomes $t^{*}_{LF}/t$ for 
single polaron hopping and 
$t^{*2}_{LF}/2 \varepsilon_p t$ if it is bipolarons which hop. 
It is for this reason that sometimes this correlation function is 
associated with some effective hopping integral $t_{eff}/2t$ which 
also in the limit $\alpha \Rightarrow 0$  is physically meaningfull
since it tends to the free electron value equal to $\frac{1}{2}$. Apart 
from these extreme limits, we shall see below, the interpretation of this correlation function in terms of an effective hopping integral has to be modified. 
Evaluating this correlation function for the two-site polaron problem 
by exact diagonalization of the Hamiltonian Eq.(\ref H) we notice that, 
as a function of $\alpha$ and $\gamma$,
the exact result for $E_{kin}/2t$ in general strongly deviates from the extreme 
polaronic LF limiting behavior. We can hence not expect that  the  
LF approximation is applicable even in conjunction with perturbative 
corrections in terms of $1/\lambda$.

In Fig.(1a) we plot $t_{eff}/2t$
as a function of $\alpha$ for different adiabaticity parameters 
$\gamma$.
At a first glance we get the impression that upon going towards 
the extreme anti-adiabatic limit i.e. $\gamma \rightarrow 0$ and 
for a fixed value of $\alpha$ we approach the extreme polaronic 
LF behavior. Yet
as can bee seen from Fig.(1b) and contrary to a well established belief 
on this matter, the LF result is approached slower and slower for a 
fixed $\gamma \ll 1$ as $\alpha$ gets bigger and bigger.These discrepancies 
with the LF results are not 
only quantitative but are qualitative in nature as shows the  $\gamma$  dependence of $t_{eff}/2t$.
The reason for this qualitatively different behavior lies in the 
LF approach itself which considers the electrons to be in localized 
states of the form given by Eq.(\ref{LFstates}) in which practically 
the entire charge of the electron and entire deformation of the 
molecules remains restricted to the molecular site on which the electron 
sits. This leads to  
$t_{eff}/t = t^{*}_{LF}/t = {\phantom{\rangle}}_0\langle\Phi(X-X_0)|
\Phi(X+X_0)\rangle_0$  at zero temperature which noticeably 
differs  from

\begin{equation}
t_{eff}/t= {\phantom{\rangle}}_0^{+}\langle\Psi(X)|\Psi(X)
\rangle_0^{-}
\end{equation} 
where $|\Psi(X)\rangle_m^{\pm}$ denote the exact eigenstates which 
replace
$|\Phi(X \pm X^0)\rangle_m$ in the expressions for the LF approximated eigenstates given by Eq.(\ref{LFstates}). As can be seen from Fig.(2) 
the exact eigenfunctions $\Psi_m^{\pm}(X)$, corresponding to the exact eigenstates
$|\Psi(X) \rangle_m^{\pm}$, differs significantly from the LF approximated 
eigenfunctions $\Phi_m^{\pm}(X)$ which for the groundstate becomes

\begin{equation}
\Phi^{\pm}_0(X) = \left( \frac{M\omega_0}{\hbar \pi}\right)^{\frac{1}{4}}
exp[-(X \pm X_0)^2(M\omega_0/\hbar)]
\end{equation}
In contrast to $\Phi^{\pm}_0(X)$ the exact eigenfunctions 
$\Psi^{\pm}_0(X)$ show a substantial deformation of the molecule 
adjacent to the one where the electron actually sits. This is born 
out in the smaller of the two peaks of $\Psi_0^{-}(X)$ for negative 
values of $\xi \equiv X\sqrt{M\omega_0/\hbar} \simeq -X_0\sqrt{M\omega_0/\hbar}$. It is the presence of this second peak 
which gives rise to a substantial increase in the value of  $t_{eff}/t$ 
over the LF approximated $t^{*}_{LF}/t$. As the coupling strength 
increases the value of $X_0$ increases roughly linearly  with $\alpha$. 
This leads to a  separation of the two main peaks of $\Psi^{+}_0(X)$ 
and $\Psi^{-}_0(X)$ respectively,
which results in an exponentially small overlap of them. On the 
contrary the overlap of the main peak of $\Psi^{+}_0(X)$ with the 
secondary peak of $\Psi^{-}_0(X)$ is of order unity since the weight 
of the secondary peak depends only weakly on the value of $\alpha$. 
Since this second contribution to $t_{eff}/t$ always remains much bigger 
than the first one, we can thus never hope to recuperate the LF result 
$t^{*}_{LF}/t$ for whatever fixed value of $\gamma < 1$.

The behavior of the oscillator wavefunction  expresses a dynamical 
delocalization of the polaronic state which has been realized a long 
time ago and for which variational calculations have given rather 
satisfactory results for the groundstate (see for instance 
ref\cite{Sho-73}). From the above study of  $t_{eff}/2t$
and of the oscillator wave function $\Psi_0^{\pm}(X)$ it becomes 
clear that in general effects of dynamical delocalization of the 
electron cannot be obtained by perturbative expansions in terms of 
$1/\lambda$ around the LF approximated oscillator wave function, 
even in the extreme antiadiabatic limit $\gamma \ll 1$. Our findings 
suggest that
it is energetically favorable for the polaron to be partially 
delocalized and  to transfer its charge by multiple hopping processes
because it is that which gives rise to the monotonically increasing 
behavior of $t_{eff}/t$
as a function of $\gamma$. These multiple hopping processes play, as 
we shall see in section 5, an important role in the correlated charge - deformation dynamics of the polaron problem.

As concerns  $t_{eff}/t$ for the 
two electron two-site system we notice from Fig.(3a) a behavior 
similar to that for the one electron problem. However as can be 
seen from Fig.(3b), in contrast to the one electron problem, for 
two electrons it varies linearly with  $\gamma$ which is an indication 
that it is bipolarons rather than polarons which hop. As $\gamma$ 
increases,  $t_{eff}/t$ tends towards a constant equal to $\frac{1}{2}$
 which is indicative of uncorrelated 
hopping of the two electrons in the system. There is a significant 
difference in slope of this correlation function calculated exactly 
and of that determined by the LF approach. Again, for reasonable 
parameters where small bipolarons are stable (such as $\alpha=1.2$ 
and $\gamma \sim 1$) we find that the exactly calculated correlaton 
function is orders of magnitude bigger than its LF expression for 
the same parameters.

In concluding this section we want to point out that although
polarons as well as 
bi-polarons exist over a large regime of parameters $\alpha$ and 
$\gamma$, their description in terms of the standard LF 
approach is generally invalid not only in the physically interesting 
regime (which lies outside the extreme strong coupling antiadiabatic 
limit $\gamma \ll 1$) but particularly in that limit, for which it 
largely overestimates the degree of self-trapping.
The failure of calculating $t_{eff}/2t$ qualitatively correctly within
the LF approach was already recognized by Fehske et al.\cite{Fehske-95} 
who attributed 
it to the zero-phonon approximation, inherent in the 
substitution Eq.(10). They showed\cite{Wellein-96} that the distribution 
of the weights 
of the $N$ phonon state in the ground state shift its maximum to higher 
values of $N$ with increasing values of $\varepsilon_p/t =2\alpha^2\gamma$. 
For a fixed values of $\alpha$ the maximum of this distribution shifts 
to higher and higher value of $N$  as $\gamma$ decreases, thus rendering the zero-phonon approximation
less and less justified. In order to visualize this behavior we plot this distribution 
$P(n)=\sum_{N_X=0}^N
(|\alpha^{+}_{0N_X}|^2+|\alpha^{-}_{0N_X}|^2)|\beta_{0N-N_X}|^2$
as a function of $N$ for a set of values of $\alpha$ (Fig.4(a)) and of
$\gamma$ (fig.4(b)). From Fig.4(b)  we can see from a 
comparison of the LF approximated distribution with the exact one for a particular choice of $\alpha=1.2$ how the LF aproach overestimates the
the self-trapping, with a maximum in $P(N)$ occuring at a significantly 
higher value of $N$ than is the case in the exactly calculated $P(N)$.  

The findings of this section moreover suggest that the effect of 
dynamical delocalization of the charge carriers in polaronic systems 
leads to a strengthening the phase 
coherence of polaronic states. If this is so, then we expect that 
for a given set of parameters $\alpha$ and $\gamma$, states described 
by the LF approach become unstable, i.e., have their spatial phase 
coherence destroyed as the temperature increases, while the exact 
states will maintain this phase coherence up to much higher 
temperatures. The key to this question lies in the wave vector 
dependence of the one particle spectral function which will be 
discussed in the next section and of the occupation number
$n_{\sigma}(k)= \langle c^{\dag}_{k\sigma}c^{\phantom{\dag}}_{k\sigma}
\rangle$ of the electronic charge carriers. Evaluating $n_{\sigma}(k)$ 
for the two lowest eigenstates within the LF approach, that is to say 
with respect to the two eigenstates given in Eq.(\ref{LFstates}) with 
$m=n=0$ we obtain  $n_{\sigma}(k=0)_{+}= n_{\sigma}(k=\pi)_{-}=\frac{1}{2}(1+exp(-\alpha^2))$ and $n_{\sigma}
(k=\pi)_{-}= n_{\sigma}(k=0)_{+}=\frac{1}{2}(1-exp(-\alpha^2))$  
respectively. This inversion of the occupation numbers between the 
ground state and the first exited state leads to the result that at 
low temperatures (bigger than the difference in energy of the two 
lowest levels but small compared to the phonon frequeny $\omega_0$)  
$n(k)= \frac{1}{2}$, independent on the wavevector $k$ and thus 
identical to the result for localized polarons, i.e. for $t=0$. 
Evaluating 
$n_{\sigma}(k=0,\pi)_{\pm}$ exactly we see from Fig.(5 ) that the 
exact result differs from that of the LF approach qualitatively 
with  $n_{\sigma}(k=0)_{\pm}$ not only being larger than $n_{\sigma}(k=\pi)_{\pm}$ for the ground state but also for the  
first excited state, provided $\alpha>\alpha_{cr}(\gamma)$ with $\alpha_{cr}(\gamma)$ increasing monotonically as $\gamma$ decreases.
$\alpha_{cr}(\gamma)$ is determined by that value of $\alpha$ for which
$t_{eff}/t$ in Fig. 1(b) tends to the saturation value $\frac{1}{2}$,
given by the free electron limit, for a given value of $\gamma$.
It is this fact which gives the polaron a stronger dynamical coherence 
than what we would expect on the basis of the LF approach and which 
shows up in the temperature variation of the effective polaron 
hopping integral $t_{eff}$, illustrated in Fig.(6). We notice that 
in the weak coupling regime $(\alpha \ll 1)$, where the electrons 
behave as quasi-free charge carriers, $t_{eff}$  decreases with 
increasing temperature. This is precisely what we expect if the 
phase coherence of the electron is destroyed by thermal fluctuations. 
On the contrary, in the polaronic regime $(\alpha \geq 1)$ we notice 
an initial small increase of $t_{eff}$ with increasing temperature
 (corresponding to a decrease in the kinetic energy of the electrons!) 
which suggests that the dynamical coherence of the polaron increases 
with increasing temperature. This holds true for temperatures up to some characteristic temperature above which this coherence is definitely 
destroyed resulting in a decrease of $t_{eff}$ with increasing 
temperature. This once more strongly contrasts with the result 
obtained within a LF approach which treats the polarons in terms 
of band states and which consequently leads to the well known 
reduction of $t^{*}_{LF}=exp(-2\alpha^2 coth(\beta \omega_0/2))$ 
with increasing temperature
(see Fig.(6)). 

These features are of course quite general and will remain when, 
instead of studying the two-site polaron problem, one deals with 
the polaron problem in an infinite lattice of finite dimension. 
There the LF approach is in fact known to give rise to results 
which contradict our findings for the two-site polaron problem 
such as the dependence of the occupation number on the wavevector 
which, apart from a small step at the Fermi vector, is a flat 
function throughout the Brillouin zone for zero temperature 
\cite{Ranninger-93}. For any realistic finite temperature, 
which clearly would be larger than the difference in energy 
of the two lowest lying eigenstates of  the Hamiltonian Eq.(\ref{H})
(i.e  $\simeq 2texp(-2\alpha^2)$), the LF approach thus would lead 
one to expect $n_{\sigma}(k)$ to be equal to $\frac{1}{2}$, 
independent on $k$ and not to show the slightest anomaly at the 
Fermi wavevector. Furthermore as concerns the temperature variation 
of  $t_{eff}$ discussed above for the 
two-site polaron problem one expects also for an infinite lattice 
a mobility which increases with increasing  temperature. The 
opposite behavior is found in the classical works on that issue 
and being based on the LF $1/\lambda$ perturbative approach \cite{Firsov-75,Holstein-59}. 

An important feature in polaron physics is the crossover from small
polarons in the anti-adiabatic strong coupling limit to large polarons 
in the adiabatic strong coupling limit. This crossover has been studied 
in great detail and can be seen most clearly if $t_{eff}$ is plotted as a function of $\varepsilon_p/t = 2 \alpha^2 \omega_0/t$\cite{Wellein-96} 
rather than as here as a function of $\alpha$. It turns out that the 
crossover is rather abrupt in the adiabatic strong coupling limit and 
becomes more spread out when going to the anti-adiabatic limit.

In the following two sections we shall discuss finer details of 
polaron dynamics which should show up in the spectral properties 
and the correlated dynamics of the charge carriers and of the 
molecular deformations which accompany them.

\section{ITINERANT VERSUS LOCALIZED POLARONS}

One of the key questions in the physics of small polarons is how to 
discern itinerant from localized polarons. In principle the answer 
to this question should be contained in the one particle density of 
states, which can be measured by photoemission experiments. 
Considering this problem within the LF approximation, the polarons 
are described by band states which are characterized by the total 
momentum $\kappa=k+q$, where $k$ denotes the momentum of the 
electron and $q$ that of the deformation which accompanies it. 
As an illustrative example for the one electron two-site polaron 
problem this implies that the LF groundstate $|LF\rangle^{+}_{00} 
\equiv |LF\rangle_{\kappa=0}$, given by Eq.(\ref{LFstates}), can 
be written as

\begin{equation}
|LF\rangle_{\kappa=0}=\frac{1}{\sqrt 2}\left(c^{\dagger}_0|
\Phi(X)\rangle_0+  c^{\dagger}_{\pi}|\Phi(X)\rangle_{\pi}\right)|
\Phi(Y-Y_0)\rangle_0
\end{equation}
where

\begin{equation}
c^{\dagger}_{(0,\pi)\sigma}=\frac{1}{\sqrt 2}\left(c^{\dagger}_{1\sigma} 
\pm c^{\dagger}_{2\sigma}\right), \;\;\;\;\;\; 
|\Phi(X)\rangle_{0,\pi}=\frac{1}{\sqrt 2}\left(|\Phi(X-X_0)\rangle_0 
\pm |\Phi(X+X_0)\rangle_0\right)
\end{equation}
Hence the scattering cross-sections

\begin{eqnarray}\label{A}
A^{(2\sigma)}_{N\sigma_1,\sigma_2,...\sigma_N}(k,\omega)=\; 
Im \frac{1}{ \pi}\int d\tau \;\exp{(i \omega \tau)} \; _{N}\langle c^{\dagger}_{k\sigma}(\tau)c^{\phantom{\dagger}}_{k\sigma}(0)\rangle_{N}
\theta(\tau) \nonumber \\
A^{(1\sigma)}_{N\sigma_1,\sigma_2,...\sigma_N}(k,\omega)= 
Im \frac{1}{\pi}\int d\tau\;\exp{(i \omega \tau)} \; _{N}\langle c^{\phantom{\dagger}}_{k\sigma}(\tau)c^{\dagger}_{k\sigma}(0)\rangle_{N}
\theta(\tau)
\end{eqnarray}
measure the intensity of emitted and absorbed electrons with momentum 
$k$ and spin $\sigma$
in a polaronic state containing $N$ electrons with spins ${\sigma_1,\sigma_2,...\sigma_N}$ in the groundstate.

Let us first consider the case of an isolated polaronic center with one 
and respectively two electrons present. We shall from now on consider only the case $U=0$. The two cross-sections, given in Eq.(\ref{A}), are then exactly  determined by

\begin{eqnarray}\label{Aloc}
A^{(1\uparrow,2\uparrow)}_{\phantom{(}0\phantom{\uparrow},1\uparrow}
(\omega)
= \sum_{l=0}^{\infty}exp(-\alpha^2)\frac{\alpha^{2l}}{l!} 
\delta(\omega-\varepsilon_p \mp l\omega_0) \nonumber \\
A^{(1\downarrow,2\downarrow)}_{\phantom{(}1\uparrow,2\uparrow\downarrow}
(\omega)= \sum_{l=0}^{\infty}exp(-\alpha^2)\frac{\alpha^{2l}}{l!}
\delta(\omega-3\varepsilon_p \mp l\omega_0)
\end{eqnarray}
which gives the well known Poisson distribution of the phonon modes 
locked together in the construction of the coherent Glauber states 
which define localized polarons. We shall now show that the overall 
features of these spectral functions
are preserved when considering a system of itinerant polarons. For 
that purpose let us consider the spectral function for electron 
emission from a one polaron ground state in the two-site polaron 
problem within the LF approach. We then obtain

\begin{eqnarray}\label{Aitn}
A^{(2\uparrow)}_{\phantom{(}1\uparrow}(k=0,\omega)_0 = 
\sum_{l\; even}^{\infty}
 \sum_{l'}^{\infty}exp(-2\alpha^2)
\frac{\alpha^{2(l+l')}}{l!l'!} 
\delta(\omega-\varepsilon_p - (l+l')\omega_0) \nonumber \\
A^{(2\uparrow)}_{\phantom{(}1\uparrow}(k=\pi,\omega)_0 = \sum_{l \; odd}^{\infty} \sum_{l'}^{\infty}
exp(-2\alpha^2)\frac{\alpha^{2(l+l')}}{l!l'!}
\delta(\omega-\varepsilon_p - (l+l')\omega_0)
\end{eqnarray}
if the system is in the ground state and 
$A^{(2\uparrow)}_{\phantom{(}1\uparrow} (k=0,\omega)_1 \; 
(= A^{(2\uparrow)}_{\phantom{(}1\uparrow}(k=\pi,\omega)_0)$ 
and $A^{(2\uparrow)}_{\phantom{(}1\uparrow} (k=\pi,\omega)_1 \; 
(= A^{(2\uparrow)}_{\phantom{(}1\uparrow}(k=0,\omega)_0)$ if it 
is in the first excited state. This shows that the bulk of the 
spectrum of itinerant polarons (i.e. $t\not=0$) is identical to 
that of localized ones except for the lowest energy part of the 
spectrum i.e. for small values of $l$. For that part of the 
spectrum the spectral weight for $k=0$ in the ground state 
equals $\simeq exp-(2\alpha^2)$ while it is identicaly zero 
for $k=\pi$. The opposite is true for the first excited state. 
These are precisely the features  expected for quasi-particles 
whose weights for their coherent part is strongly reduced. The 
exact results of the spectral functions bare this out and in 
Fig.(7a) we present $A^{(2\uparrow)}_{\phantom{(}1\uparrow}
(k=0,\omega)_0$ for values of $\gamma$ and $\alpha$ which 
characterize well defined polaronic states. In order to compare $A^{(2\uparrow)}_{\phantom{(}1\uparrow}(k=0,\omega)$ with $A^{(2\uparrow)}_{\phantom{(}1\uparrow}(\omega)$ for 
localized polarons (i.e. $t=0$) as well as  with $A^{(2\uparrow)}_{\phantom{(}1\uparrow}(k=\pi,\omega)$ we do 
this for finite rather than zero temperature since for all intents 
and purposes the level splitting between the fist two lowest energy levels (corresponding to the bandwidth  $2t^*$ of the coherent polaron motion) 
is extremely small, since under normal 
conditions we are dealing with temperatures $T$ such that 
$2t^* \ll k_BT \ll \omega_0$. Under those circumstances the 
LF approach yields scattering cross-sections $A^{(2\uparrow)}_{\phantom{(}1\uparrow}(k,\omega)$ which become 
wave vector independent and thus undistinguishable from purely 
localized polarons. The exactly calculated spectral functions 
become, up to a $k$ dependent scaling factor, practically 
identical with $A(\omega)$ for the entire frequency regime 
except for the low frequency part, where however the spectral weight 
is extremely small (see Fig.7b). These findings suggest that small polarons
might never exist in form of coherent Bloch like states, a feature which is supported by the dynamics of the polaron motion, which will be discussed in section V.

One of the prime problems in the physics of the many polaron problem 
is the question of how to discern a system of essentially 
non-interacting polarons from one in which they sense a strong 
attraction between 
them and which ultimately leads to the formation of tightly bound 
pairs (bipolarons). In principle this question  can be resolved by 
photoelectron spectroscopy which measures  the one particle density 
of states for polarons. In fact what is required in order 
to discriminate between a many body ground state containing essentially 
unpaired from one containing essentially paired polarons is both, 
photoemission and inverse photoemission spectroscopy. Testing the 
system with solely photoemisson we can not decide about the nature 
of the many body ground state. This can be illustrated on the basis 
of the two-site polaron system 
containing one and respectively two electrons. The scattering cross 
section looks essentially similar for the two cases. In Fig.(8,a-d) 
we plot for different 
values of coupling strength $\alpha$ the photoemission spectrum $A^{(2\uparrow{\phantom{\uparrow}})}_{\phantom{(}2\uparrow\downarrow}
(k=0,\omega)$ where an electron is emitted out of a two-electron two-site polaron system. This spectrum has the same 
form as  $A^{(2\uparrow)}_{\phantom{(}1\uparrow}(k=0,\omega)$ 
(see Fig.(7a)) which corresponds to the situation of an electron emitted 
out of a one-electron two-site system.  On the contrary, the inverse photoemission 
spectrum can clearly distinguish if the final state corresponds to 
a state with essentially uncoupled polarons or polaron pairs. 
Looking at this spectral function for  the one-electron two-site 
system i.e., $A^{(1\downarrow)}_{\phantom{(}1\uparrow}(k=0,\omega)$  
we see from Fig.(9a-d) that, as we increase the coupling strength, 
a two peak structure emerges. The energy gap which separates this 
two peak structure amounts to   $2\varepsilon_p$, which represents 
precisely the binding energy between two polarons in the strong 
coupling limit. The low energy peak of this spectrum corresponds 
to a final state given by a biporaron in its bonding singlet state, 
having energy $\simeq 4\varepsilon_p$. The high energy peak arises 
from a bipolaronic state in its antibonding singlet and respectively 
triplet state having an energy $\simeq 2\varepsilon_p$\cite{comment}. 
The difference in energy  between these two contributions in the 
spectral function for inverse photoemission is hence just the binding 
energy 
of a bipolaron and thus can serve as a signature for a bipolaronic 
many polaron ground state.

It is even moreover illustrative to study the one particle density 
of states for those situations.
In Figs.(10,a-d) we plot the density of states per spin for the one 
electron two-site system at zero temperature 
\begin{equation}
\rho_1(\omega) = \sum_{k=0,\pi} \left( A^{(2\uparrow)}_{\phantom{(}1\uparrow}(k,\omega)_0 + A^{(1\uparrow)}_{\phantom{(}1\uparrow}(k,\omega)_0 \right)
\end{equation}
as a function of coupling strength $\alpha$ and  for a given fixed 
value of adiabaticity parameter $\gamma= 1.1$. For small values of 
$\alpha$ the density of states is characterized by two peaks centered 
around energies $\omega \simeq 0$ and $\omega \simeq 2t$, which in 
essence represents quasi free electrons in the bonding and 
respectively anti-bonding states for this small system. As $\alpha$ 
increases these two peaks spread and eventually evolve into two well 
separated peaks. This density of states is similar to that of the 
Hubbard model in the dilute limit (i.e., far away from half filling) 
which shows indications for the presence of an upper and a lower 
Hubbard band, separated by the Hubbard $U$ repulsion energy, but 
without any clear gap between those two bands.
Similarly to the Hubbard problem, the two peaks in the polaron 
problem are separated by the energy of attraction between two 
polarons, which in case of strong coupling (such as illustrated 
in Fig.(9d) for $\alpha = 2.2$) amounts to $2\varepsilon_p$.

A similar density of states per spin is obtained  for the two 
electron two-site system
\begin{equation}
\rho_2(\omega) = \sum_{k=0,\pi}\left( A^{(2\uparrow)}_{\phantom{(}2\uparrow\downarrow}(k,\omega)_0 + A^{(1\uparrow)}_{\phantom{(}2\uparrow\downarrow}(k,\omega)_0  
\right)
\end{equation}
which we illustrate in Figs.11(a-d) as a function of coupling 
strength $\alpha$ and for a given fixed value of adiabaticity 
parameter $\gamma = 1.1$.
We again see a similar evolution of the density of states as we 
increase $\alpha$ but now the separation in energy is equal to 
$4\varepsilon_p$, which can be clearly distinguished from the case 
representative for uncoupled polarons (the one electron two site 
problem considered above).
A clear energy gap given by $2 \varepsilon_p$ now appears separating 
the low and high frequency peak structures. Integrating $\rho_2(\omega)$ 
up to some value
$\omega= \varepsilon_F$ such that it becomes equal to 2 (which 
corresponds to the half filled band case and to two electrons in 
our system) we find that $\varepsilon_F$ lies precisely in the 
middle of the gap of the density of states.
This again is reminiscent of the problem of strongly correlated 
electrons for the half filled band Hubbard model. These features 
examined here for the two-site polaron system are expected to 
hold true generally for any interacting many polaron system on 
a lattice.

\section{Correlated charge-deformation dynamics}

The polaron problem presents a highly non-linear dynamical system 
in which the charge and deformation fluctuations are intricately 
coupled together. This leads to a dynamics of the molecular 
deformations being driven by the dynamics of the charge carriers 
in the strong coupling anti-adiabatic regime ($\alpha \gg 1, 
\gamma \ll 1)$ . On the contrary it is the dynamics of the molecular deformations which drives the dynamics of the charge carriers in 
the weak coupling adiabatic regime ($\alpha \ll 1, 
\gamma \gg 1)$. In general this leads to a dynamics for the 
charge and of the molecular deformations which is composed of a 
common slow oscillation and fast ones superposed on it. The fast 
oscillations for the charge dynamics have a frequency 
$\widetilde{t}$ of the order of $t$, while those for the 
deformation dynamics have a frequency $\widetilde{\omega}_0$
which is of the order of $\omega_0$.

In this section we shall study the evolution of those dynamical 
properties of small polarons when we go from the strong coupling 
anti-adiabatic regime to the strong coupling adiabatic one, with a 
special emphasis on the behavior  in the crossover regime. In order 
to illustrate this behavior we evaluate the time dependent 
correlation functions for the charge redistribution and molecular 
deformations:

\begin{equation}
\chi_{nn}(\tau)=\langle (n_{1\sigma}(\tau)-n_{2\sigma}(\tau))
(n_{1\sigma}(0)-n_{2\sigma}(0)) \rangle, \;\;\;\;  
\chi_{xx}(\tau)=\langle X(\tau)X(0) \rangle
\end{equation}
In Figs.12(a-d) we plot these correlation functions (normalized with 
respect to their $\tau=0$ values) as a function of time $\tau$ in 
units of $\omega_0$ for different values of the $\gamma$ and fixed 
$\alpha$. We notice that the charge dynamics qualitatively tracks globally
the 
behavior expected on the basis of the LF approximation in the 
anti-adiabatic limit (i.e., for $\gamma = 0.1$, Fig.12(a)) but with 
superposed small amplitude fast charge oscillations wih a frequency $\widetilde{t}$ which is large compared to the unrenormalized 
electron hopping integral $t$. The 
dynamics of the molecular deformation follows in a coherent fashion 
that of the charge and exhibits superposed molecular vibrations with 
a frequency $\widetilde{\omega}_0 \simeq \omega_0$. As we increase 
$\gamma$ (i.e., for $\gamma = 1.1$, Fig.12(b)), the amplitude of the 
fast charge oscillations become increasingly important and their 
oscillation frequency $\widetilde{t}$  decreases. The opposite 
behavior is obtained for the molecular deformation oscillations, 
whose frequency  $\widetilde{\omega}_0$ of the fast oscillatory 
behavior increases while the corresponding amplitude diminishes. 
The system is then no longer described by the LF approach (as we can see 
from the comparison made for this case) and the charge dynamics is 
now controlled by multiple hopping processes with concomitantly 
reduced amplitude fluctuations of the molecular deformations.
Upon further increasing $\gamma$ we arrive at a situation where the 
frequencies and amplitudes of those two dynamical variables, 
characterizing the charge and deformation, become comparable 
to each other i.e., for $2\widetilde{t} \simeq \widetilde{\omega}_0$ 
and  we 
enter the crossover regime (i.e., for $\gamma = 1.6$,  
Fig.12(c)).
This regime is characterized 
by a temporarily alternating behavior between essentially 
self-trapped anti-adiabatic small polarons (manifest in a 
substantial reduction of the amplitude fluctuations of the 
charge dynamics) and a behavior reminiscent of itinerant 
adiabatic polarons as seen upon a further increase of 
$\gamma$ (ie.,$\gamma = 2.0$, Fig.12(d)). Such fluctuations 
in the amplitudes of the charge occur over a 
time scale which is large compared to the inverse hopping rate  
$\widetilde{t}$ and is accompanied by phase slips in the fast 
oscillatory behavior of the charge as well as the deformation 
fluctuations.

The evolution of the correlation functions for the 
charge and deformation fluctuations of the two-electron two-site problem
as a function of the adiabaticity 
parameter $\gamma$ and fixed coupling constant $\alpha$ follows a similar behavior to that of the one-electron two-site problem, as shown in 
Figs.13(a-d). Again we can identify a crossover between small 
bi-polarons and essentially uncorrelated two electrons which 
occurs for $\gamma \simeq 1.3$ for the particular choice $\alpha=0.6$
 

This crossover between self-trapped polarons, respectively bi-polarons,
and quasifree electrons in the phase space of $\alpha$ and $\gamma$ 
corresponds to the characteristic value of $\gamma$ where the 
kinetic energy of the electrons 
$E_{kin}=-2t\langle c_{1,\uparrow}^{\dagger} c_{2,\uparrow} \rangle$ 
approaches its maximal value of the free electron limit, 
while the potential energy of the electrons 
$E_{pot} = -2\lambda \langle (n_{1,\uparrow}+
n_{1,\downarrow}) u_1\rangle$ 
tends to its minimum value 
$-\frac{1}{\sqrt2}\lambda Y_0 \langle (n_{1,\uparrow}+n_{1,\downarrow}) \rangle$,
obtained in the limit $\gamma \Rightarrow \infty$. 
This can be seen from the behavior of $E_{kin}$ illustrated in Fig.1(b) 
and of $E_{pot}$ depicted in 
Fig.14 for $\alpha=1.2$ and $\gamma\simeq 1.6$ for the one-electron 
two-site problem. The crossover between self-trapped small polarons and 
quasi-free electrons thus appears to be driven by a competition between 
the kinetic and potential energy of the electrons; the first one 
favoring a delocalization of them while the second one inciting them to 
localize on the molecular sites. In an infinite solid state system such 
a scenario would suggest a quantum phase transition between a metal and a polaronic insulator as  proposed a long time ago by Landau and Froehlich
\cite{Landau-33}. For systems with low carrier concentrations such 
localized polarons have been verified experimentally in metal halide 
where optically excited excitons get localized\cite{Ueta-86}. There  
is at present no exact theorem as to whether pure electron 
phonon-systems can show such a polaronic insulator. The present  
exact proofs against that\cite{exactproves} 
hinge on suppositions of the phonon spectra which may not be realistic 
for real materials.

We finally should like to point point out that no significant 
changes in the phonon distribution of the displaced oscillators are 
observed when going through this crossover regime. This can be verified 
from the plot in  Fig.4(b) of $P(N)$ for a fixed $\alpha=1.2$ and
upon varying $\gamma$.

In Table I we 
summarize our findings on the charge-deformation dynamics of a polaronic 
system. We compare for that purpose the renormalized vibrational frequency $\widetilde{\omega}_0$, the renormalized
electron hopping integral  $\widetilde{t}$, the kinetic energy of the 
electrons $t_{eff}/t$ 
and the physical charge transfer 
rate $t^*$ for different values of the 
adiabaticity parameter $\gamma$ and for a fixed coupling constant 
$\alpha$. In the limit of strong adiabaticity, $t^*$ tends to the LF
value $t^*_{LF}$ and follows as a function of $\gamma$ the behavior of 
$(E_1-E_0)/2$ which denotes the difference in energy of the two lowest eigenstates.
We also indicate the spectral weight $Z$ of the lowest energy contribution 
to the scattering cross-section and notice that for the anti-adiabatic 
limit it scales with $t^*/t$. This is an indication that the low 
frequency part 
of the scattering cross-section corresponds to coherent states but with a spectral weight which is extremely small. It is presently not clear 
whether such weak 
coherent features, characteristic of itinerant small polarons, will
persist if one treats the polaron problem on a infinite lattice. 
Calculations based on infinite dimensions\cite{Ciuchi-95} which 
show such itinerant 
behavior do neglect totally the frequency renormalization of the 
vibrational motion of the atoms, which we consider as the prime 
cause for dephasing of the correlated charge-deformation dynamics 
and ultimately believed to be responsible for the destruction of 
itinerant polaronic states.

The strong coupling between the charge and the molecular deformations 
thus  manifests itself not only in a strong renormalization of the 
molecular vibrational frequency $\omega_0$ becoming 
$\widetilde{\omega}_0$ but also in a strong renormalization of 
the intrinsic hopping integral
$t$ renormalized into $\widetilde{t}$. As we go from the 
anti-adiabatic limit toward the adiabatic one (for fixed value 
of $\alpha$), we observe a substantial decrease in 
$\widetilde{t}/t$ and a concomitant increase in  
$\widetilde{\omega}_0$. These are effects which should be 
observable by spectroscopic measurements such as infrared or 
Raman scattering for the vibrational modes.

>From inspection of Figs.(12,13) we notice the sizeable dynamical 
delocalization of the polaron and bi-polaron respectively
 as we approach the crossover regime. 
In the crossover regime itself
this delocalization alternates between partly quasi-static 
delocalization, suggesting almost localized yet extended polaron 
states and dynamical delocalization reminiscent of almost free 
carriers, which however remain dynamically tied to a given 
molecule for some appreciable time.
To be more specific, for the particular case illustrated in Fig.12(c), 
we find for the period of time when a 
quasi-static polaron is stable, a charge distribution given by: 
$\langle n_{1,\sigma}\rangle \simeq 0.75$ and $\langle n_{2,\sigma}
\rangle \simeq 0.25$. On the other hand for the dynamically 
delocalized polaron we find that the charge distribution fluctuates 
over a characteristic time given by $\widetilde{t}$ between  $\langle n_{1,\sigma}\rangle \simeq 1.0$, $\langle n_{2,\sigma}\rangle 
\simeq 0$ and$\langle n_{1,\sigma}\rangle \simeq 0.5$, 
$\langle n_{2,\sigma}\rangle \simeq 0.5$. Such temporal 
fluctuations were initially hypothezised by us a long time 
ago\cite{Ranninger-85} which led to the Boson-Fermion model 
for intermediary coupling electron phonon systems, which may have 
some relevance for our understanding of High $T_c$ superconductors\cite{Ranninger-95,Ranninger-96}.

\section{SUMMARY}
The main objective of this work was the study of the intricate 
dynamics of the polaron problem involving the dynamical behavior 
of the charge carriers and that of the molecular deformations 
which surround them. We find that in the anti-adiabatic regime 
for small polarons the molecular deformations follow in a coherent 
fashion the redistribution of the charge, while in the adiabatic 
regime it is the charge redistribution which follows the molecular 
deformations. 
In the crossover regime between those two limiting cases we find 
that the dynamical  behavior of the polaronic charge carriers 
alternates between self-trapped polarons and almost free carrier 
behavior. The time scale over which these different behaviors are 
realized is typically an order of magnitude bigger than the 
intrinsic hopping rate i.e., of the order $10 \times 2\pi/t$. 
This crossover regime is characterized by strong renormalization 
of the intrinsic hopping integral as well as of the bare phonon 
frequency, which in this regime become equal. Phase slips in the 
fast oscillatory components of the charge and molecular deformation 
fluctuations are the result of this. Such effects are expected to be 
essential for a proper description of polaron damping,  sofar having 
been treated only within the LF approach\cite{Loos-93}, and
which is unable to account for the effects described here. 

The question of  how to distinguish itinerant polarons from localized 
ones was studied here on the basis of the one particle spectral function 
and its temperature dependence. We showed that at zero temperature 
the respective spectral functions for localized and itinerant polarons 
differ from each other only very slightly except for the low frequency 
regime of the spectral function, where they show an increased spectral 
weight  for small wave vectors if the polarons are itinerant. As the 
temperature increases this difference disappears and it is, in principle, 
no longer possible by spectroscopic means to distiguish between 
localized and itinerant small polarons. This may explain the puzzling 
results in the photoemission spectra for certain High $T_c$ 
superconducting cuprates for which a wave vector independent 
spectral function was observed in the normal state and consequently 
was interpretated as indications for localized charge carriers\cite{Margaritondo-95}. The question of a polaronic insulator 
versus a polaronic metal has been touched upon here only from the point 
of view of the single particle properties. The polaron problem is however 
presents a problem of electrons in a system with impurity centers with dynamically varying energies and thus contains features similar to those
of the Anderson localization. The relevant quantity to be studied hence 
is the conductivity.So far a few attempts in this direction have  been 
made on the basis of exact diagonalization studies in finite systems
\cite{Fehske-97} attempting to determine whether there is or not 
a finite Drude component in the optical conductivity. 

Finally, our exact diagonalization studies on the two-site molecular 
Holstein polaron model permitted us to discuss the limitations of 
the standard LF approach. Our rather unexpected  and perhaps widely  
unrecognized findings are that this approach, which is generally 
believed to become exact in the limit of anti-adiabaticity and an 
electron phonon coupling going to infinity, actually diverges most 
from the exact results precisely in this limit. The reasons for that 
can be traced back to the zero-phonon approximation inherent in the LF 
approach, based on the relation Eq. (10) and
which, with increasing coupling strength, is increasingly strongly violated\cite{Fehske-95}.

Our analysis of the various spectral functions and the density of 
states shows that the major part of the spectrum must be considered as 
being due to incoherent rather than coherent polaron dynamics; the 
latter having vanishingly
small spectral weight of order $\exp(-2\alpha^2)$. This result confirms 
our earlier findings on the many polaron problem for infinite lattices\cite{Alexandrov-92} and examinations of the single polaron 
problem in infinite dimensions\cite{Ciuchi-95}.

\section{ACKNOWLEDGEMENT}
We are indebted to S. Ciuchi, H.Fehske and J.M.Robin for many valuable discussions and J.M.R. for 
making available to us certain of his independently derived results 
on the evaluation of the spectral functions. E.V.L.M. acknowledges 
a post dostoral fellowship
from the Brazilian council for national Research $(CNP_q)$.

$\dagger$ On leave of absence from the Departamento de F\'isica, 
Universidade 
Federal Fluminense, Rio de Janeiro, Brazil.

\newpage


\begin{figure}
{\bf Fig.1} $t_{eff}/2t= -E_{kin}/2t = \langle c^{\dag}_{1\sigma}c^{\phantom{\dag}}_{2\sigma}\rangle$
for a single electron as a function of 
$\alpha$ for different adiabaticity parameters $\gamma$ for the 
two-site polaron problem (a). Comparison of $t_{eff}/t$ evaluated 
by exact diagonalization and its approximative value given by the 
LF approach as a function of $\gamma$ for several values of 
$\alpha$ (b).
\end{figure}

\begin{figure}
{\bf Fig.2} The exact oscillator wave function $\Psi_0^{+}(\xi)$ 
with $\xi=X\sqrt{M\omega_0/\hbar}$ for a single electron in the 
two-site polaron system for different values of $\alpha$. 
\end{figure}

\begin{figure}
{\bf Fig.3} $t_{eff}/t$ for two electrons as a function of $\alpha$ 
for different adiabaticity parameters $\gamma$ for the two-site
polaron problem (a). Comparison  of $t_{eff}/t$ evaluated by exact 
diagonalization with its approximated value given by the LF 
approach as a function of $\gamma$ for different values of $\alpha$ (b).
\end{figure}

\begin{figure}
{\bf Fig.4} (a) The phonon number distribution $P(N)$ for a given value of 
$\alpha=1.2$ and for various values of $\gamma$. Notice that as  $\gamma$ decreases, the distribution function tends to that of localized polarons, practically indistinguishable from that of $\gamma=0.1$.
(b)  The phonon number distribution  for a given value of 
$\gamma=1.6$ and for various values of $\alpha$. Notice the in general significant difference between the LF approximated distribution 
function and the exact results, illustrated here for $\alpha=1.2$.
\end{figure}

\begin{figure}
{\bf Fig.5} The wave vector dependence of the polaron occupation 
number $n^{pol}_{\sigma}(k)_{\pm}$ for the two lowest lying states. 
For comparison we also plot the result for the LF approximated 
states given in Eq.(\ref{LFstates}).
\end{figure}

\begin{figure}
{\bf Fig.6} $t_{eff}/t$ (normalized to its zero temperature value) 
as a function of temperature (in units of $\omega_0$) for fixed 
$\gamma = 1.1$ and for different 
values of $\alpha$ and compared with the LF approach results. 
\end{figure}

\begin{figure}
{\bf Fig.7}
The scattering cross-section $A^{(2\uparrow)}_{\phantom{(}1\uparrow}
(k=0,\omega)_0$ for  $\gamma=1.1$ and $\alpha=2.2$ and for $k_BT = 0.1 
\omega_0$ (a). Comparison of the scattering cross-sections $A^{(2\uparrow)}_{\phantom{(}1\uparrow}(k=0,\omega)_0$ and $A^{(2\uparrow)}_{\phantom{(}1\uparrow}(\pi,\omega)_0$ with each 
other as well as with that for localized polarons i.e., $A^{(2\uparrow)}_{\phantom{(}1\uparrow}(\omega)_0$ for $k_BT =0.1 
\omega_0$ and $T=0$ (b). These scattering cross-sections have been 
obtained by broadening the set of $\delta$ functions by Gaussians of 
width $\Delta\omega = 0.1 \omega_0$.
\end{figure}

\begin{figure}
{\bf Fig.8(a-d)}
The scattering cross-section $A^{(2\uparrow)}_{\phantom{(}2\uparrow\downarrow}(k=0,\omega)_0$ 
for photoemission from a two electron two-site system as a function 
of increasing electron phonon coupling $\alpha$. These scattering 
cross-sections have been obtained by 
broadening the set of $\delta$ functions by Gaussians of width 
$\Delta\omega = 0.1 \omega_0$.
\end{figure}

\begin{figure}
{\bf Fig.9}
The scattering cross-section $A^{(1\uparrow)}_{\phantom{(}1\uparrow}(k=0,
\omega)_0$ for inverse photoemission from a one-electron two-site system as 
a function of increasing electron phonon coupling $\alpha$. The scattering cross-section has been obtained by 
broadening the set of $\delta$ functions by Gaussians of 
width $\Delta\omega = 0.1 \omega_0$.
\end{figure}

\begin{figure}
{\bf Fig.10(a-d)}
The evolution of the density of states $\rho_1(\omega)$ for the 
one-electron two-site polaron problem  as the electron phonon coupling 
$\alpha$ increases and for a fixed adiabaticity parameter $\gamma = 1.1$. 
The densities of states  have been obtained by 
broadening the set of $\delta$ functions by Gaussians of width 
$\Delta\omega = 0.1 \omega_0$.
\end{figure}

\begin{figure}
{\bf Fig.11(a-d)}
The evolution of the density of states  $\rho_2(\omega)$ for the 
two-electron two-site polaron problem as the electron phonon coupling 
$\alpha$ increases and for a fixed adiabaticity parameter $\gamma = 1.1$.
The densities of states  have been obtained by 
broadening the set of $\delta$ functions by Gaussians of width 
$\Delta\omega = 0.1 \omega_0$.
\end{figure}

\begin{figure}
{\bf Fig.12a}
$\chi_{nn}$ and $\chi_{xx}(t)/\langle X^2 \rangle$ for $\alpha =1.2$ 
and $\gamma=0.1$ for the one-electron two-site polaron system.
\end{figure}

\begin{figure}
{\bf Fig.12b}
$\chi_{nn}$ and $\chi_{xx}(t)/\langle X^2 \rangle$ for $\alpha =1.2$ 
and $t\gamma=1.1$ for the one-electron two-site polaron system.
\end{figure}

\begin{figure}
{\bf Fig.12c}
$\chi_{nn}$ and $\chi_{xx}(t)/\langle X^2 \rangle$ for $\alpha =1.2$ 
and $\gamma=1.6$  for the one-electron two-site polaron system.
\end{figure}

\begin{figure}
{\bf Fig.12d}
$\chi_{nn}$ and $\chi_{xx}(t)/\langle X^2 \rangle$ for $\alpha =1.2$ 
and $\gamma=2.0$ for the one-electron two-site polaron system.
\end{figure}

\begin{figure}
{\bf Fig.13a}
$\chi_{nn}$ and $\chi_{xx}(t)/\langle X^2 \rangle$ for $\alpha =0.6$ 
and $\gamma=0.2$ for the two-electron two-site polaron system.
\end{figure}

\begin{figure}
{\bf Fig.13b}
$\chi_{nn}$ and $\chi_{xx}(t)/\langle X^2 \rangle$ for $\alpha =0.6$ 
and $\gamma=0.8$ for the two-electron two-site polaron system.
\end{figure}

\begin{figure}
{\bf Fig.13c}
$\chi_{nn}$ and $\chi_{xx}(t)/\langle X^2 \rangle$ for $\alpha =0.6$ 
and $\gamma=1.3$ for the two-electron two-site polaron system.
\end{figure}

\begin{figure}
{\bf Fig.13d}
$\chi_{nn}$ and $\chi_{xx}(t)/\langle X^2 \rangle$ for $\alpha =0.6$ 
and $\gamma=1.6$ for the two-electron two-site polaron system.
\end{figure}

\begin{figure}
{\bf Fig.14}
Comparison of the electronic kinetic and potential energy 
$E_{pot} = -2\lambda \langle (n_{1,\uparrow}+
n_{1,\downarrow}) u_1 \rangle$ ( as a function 
of $\gamma$ for a fixed $\alpha= 1.2$ for the one-electron two-site 
problem. The crossover regime between self-trapped polarons and 
quasi-free electrons occurs when the kinetic energy tends to its maximal 
value while that of the potential energy tends to its minimal value.
\end{figure}
\newpage
\begin{table}
\begin{center}
\begin{tabular}{|c|c|c|c|c|c|c|c|} \hline  
\makebox[1.5cm]{\centering $\gamma =t/\omega_0$}      &
\makebox[1.5cm]{\centering $\tilde{t}/t$} &
\makebox[1.5cm]{\centering $\tilde{\omega}_0/{\omega_0}$} &
\makebox[1.5cm]{\centering $\tilde{t}/\tilde{\omega}_0$} &
\makebox[1.5cm]{\centering $t_{eff}/t$} &
\makebox[1.5cm]{\centering $\Delta E/2 t$} &
\makebox[1.5cm]{\centering $t^*/t$} &
\makebox[1.5cm]{\centering $Z$}                                                                   \\  \hline\hline 
0.1  &   -    &  1.03  &   -   & $1.00\;10^{-1}$ & $5.65\;10^{-2}$ & $5.61\;10^{-2}$ & $6.11\;10^{-2}$  \\  \hline
0.3  &  5.66  &  1.08  &  1.57 & $1.90\;10^{-1}$ & $5.73\;10^{-2}$ & $5.66\;10^{-2}$ & $7.25\;10^{-2}$  \\  \hline
0.5  &  3.05  &  1.14  &  1.34 & $2.82\;10^{-1}$ & $5.95\;10^{-2}$ & $5.98\;10^{-2}$ & $8.59\;10^{-2}$  \\  \hline
1.1  &  1.56  &  1.42  &  1.21 & $5.64\;10^{-1}$ & $6.98\;10^{-2}$ & $6.79\;10^{-2}$ & $1.34\;10^{-1}$  \\  \hline
1.3  &  1.41  &  1.52  &  1.07 & $6.20\;10^{-1}$ & $7.63\;10^{-2}$ & $6.90\;10^{-2}$ & $1.49\;10^{-1}$  \\  \hline
1.6  &  1.25  &   -    &   -   & $7.12\;10^{-1}$ & $7.84\;10^{-2}$ & $7.85\;10^{-2}$ & $1.70\;10^{-1}$  \\  \hline
1.7  &  1.28  &   -    &   -   & $7.38\;10^{-1}$ & $7.97\;10^{-2}$ & $8.03\;10^{-2}$ & $1.75\;10^{-1}$  \\  \hline
2.0  &  1.19  &   -    &   -   & $8.00\;10^{-1}$ & $8.26\;10^{-2}$ & $8.26\;10^{-2}$ & $1.90\;10^{-1}$  \\  \hline
\end{tabular}
\end{center}
\caption{The variation of the renormalized frequency of the 
deformation oscillations  $\widetilde{\omega}_0$ and of the 
renomalized hopping rate $\widetilde{t}$ as a function of the 
adiabaticity parameter $\gamma$ for fixed coupling constant 
$\alpha=1.2$. Notice that as we approach the crossover regime, 
the time scales of these two oscillations become equal i.e., $\tilde{t}/\tilde{\omega}_0 \rightarrow 1$. We also compare the frequency $t^*$ of the slow polaronic motion with the splitting of the two lowest eigenvalues $\Delta E/2 t$} and  $t_{eff}$ (the electron kinetic energy). Notice that the spectral weight $Z$ of the lowest frequency pole of the electron Green's function
scales fairly well with the renormalization factor for the polaron bandwidth $t^*/t$.
\end{table}


\begin{references}
\bibitem{comment} This corrects our earlier results in 
ref(\cite{Ran-92})
in which certain contributions to these correlation functions 
were inadvertently omitted.
\end{references}
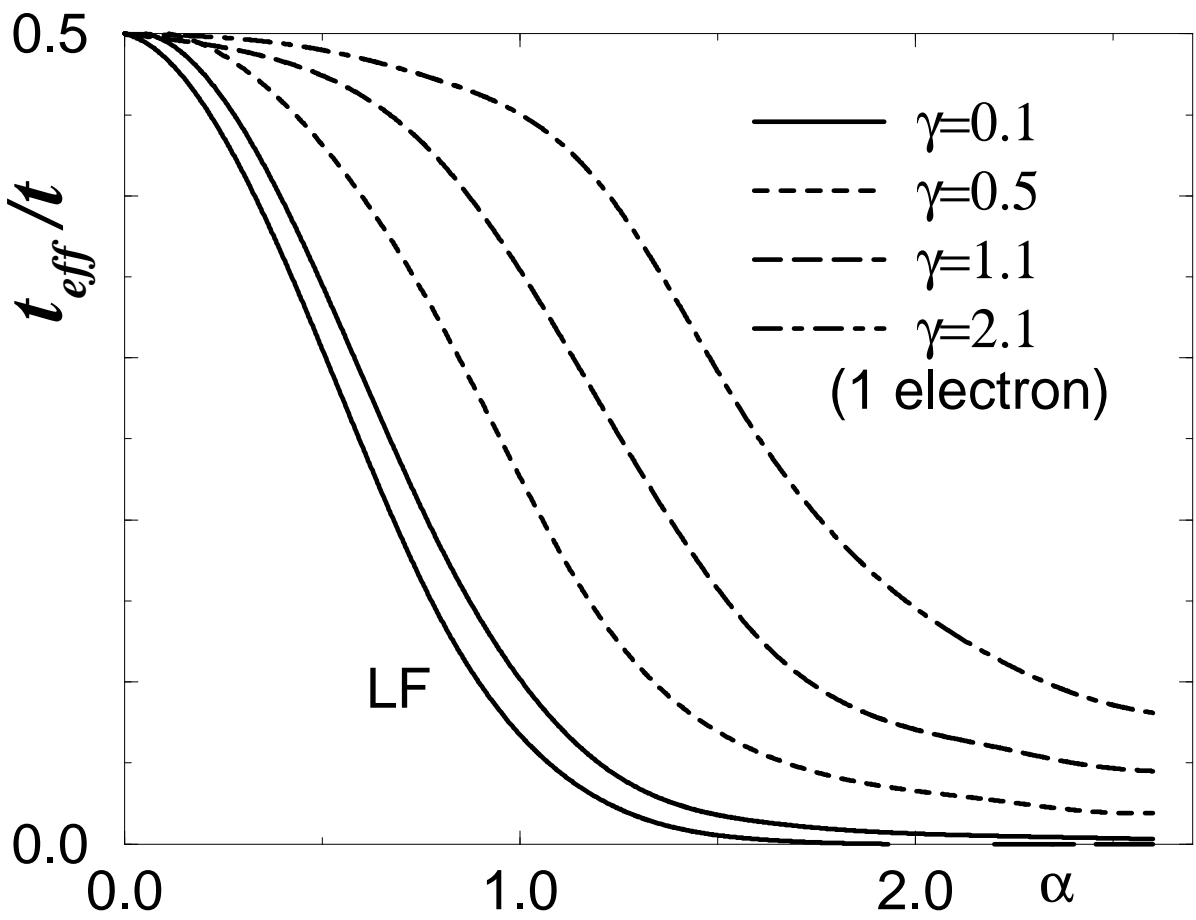
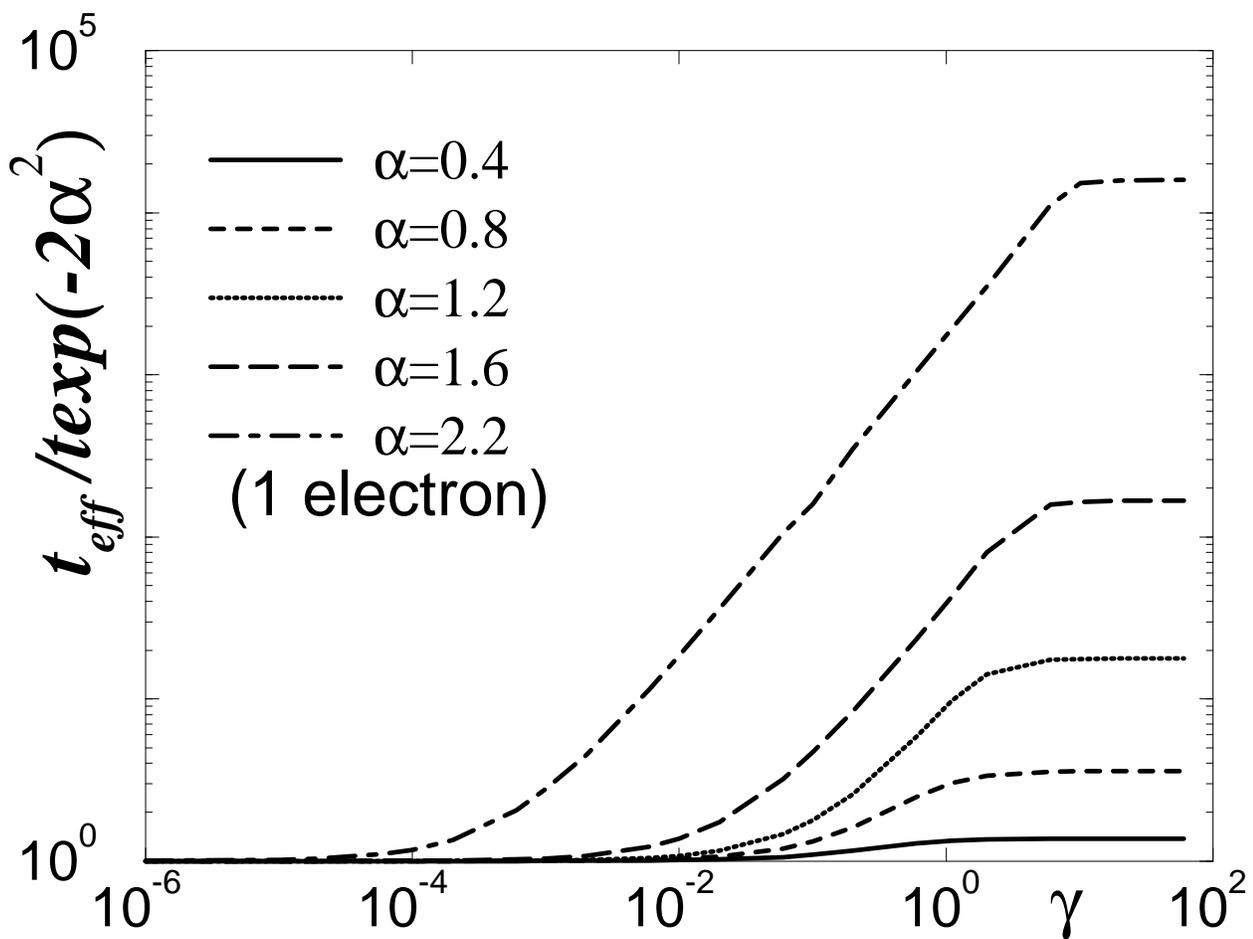
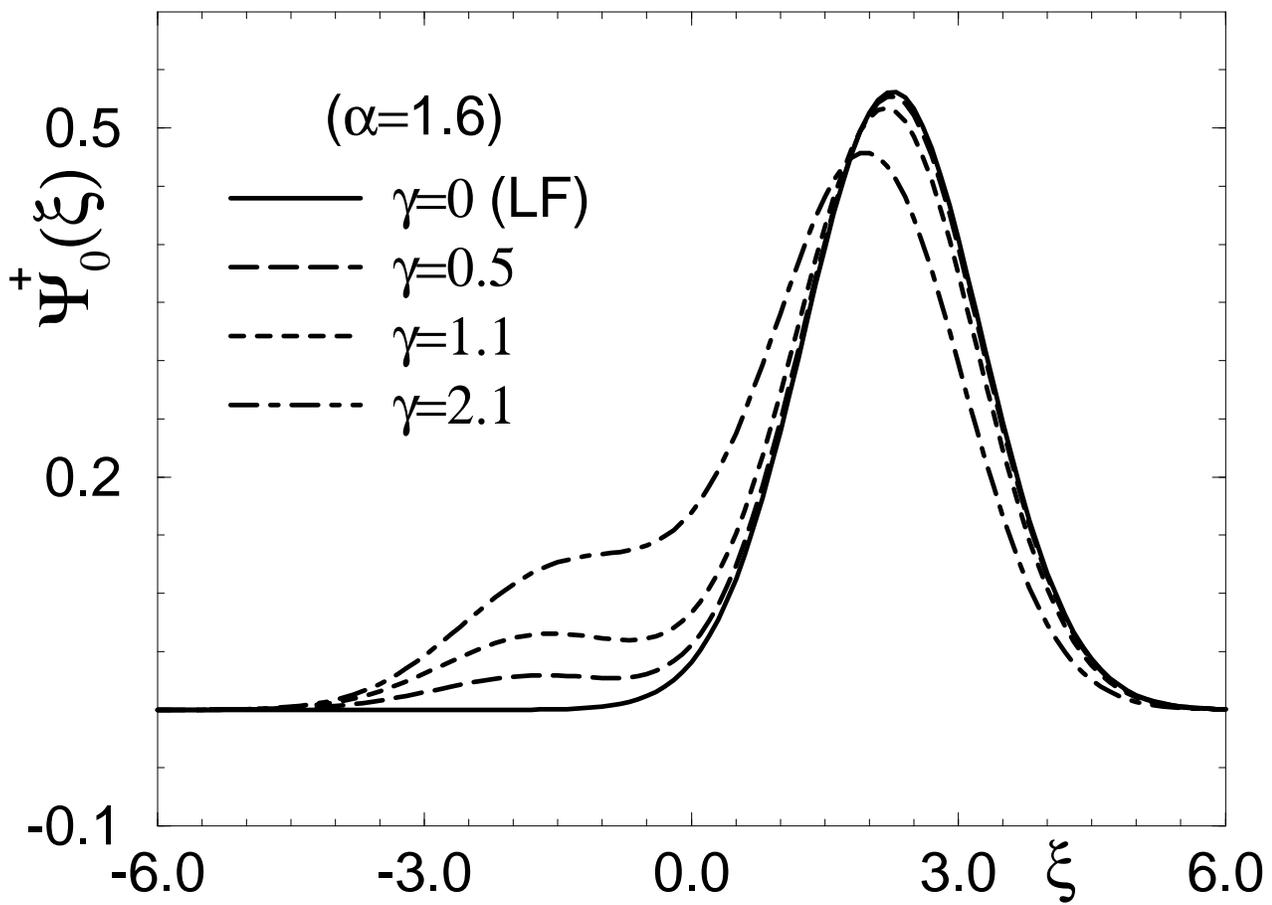
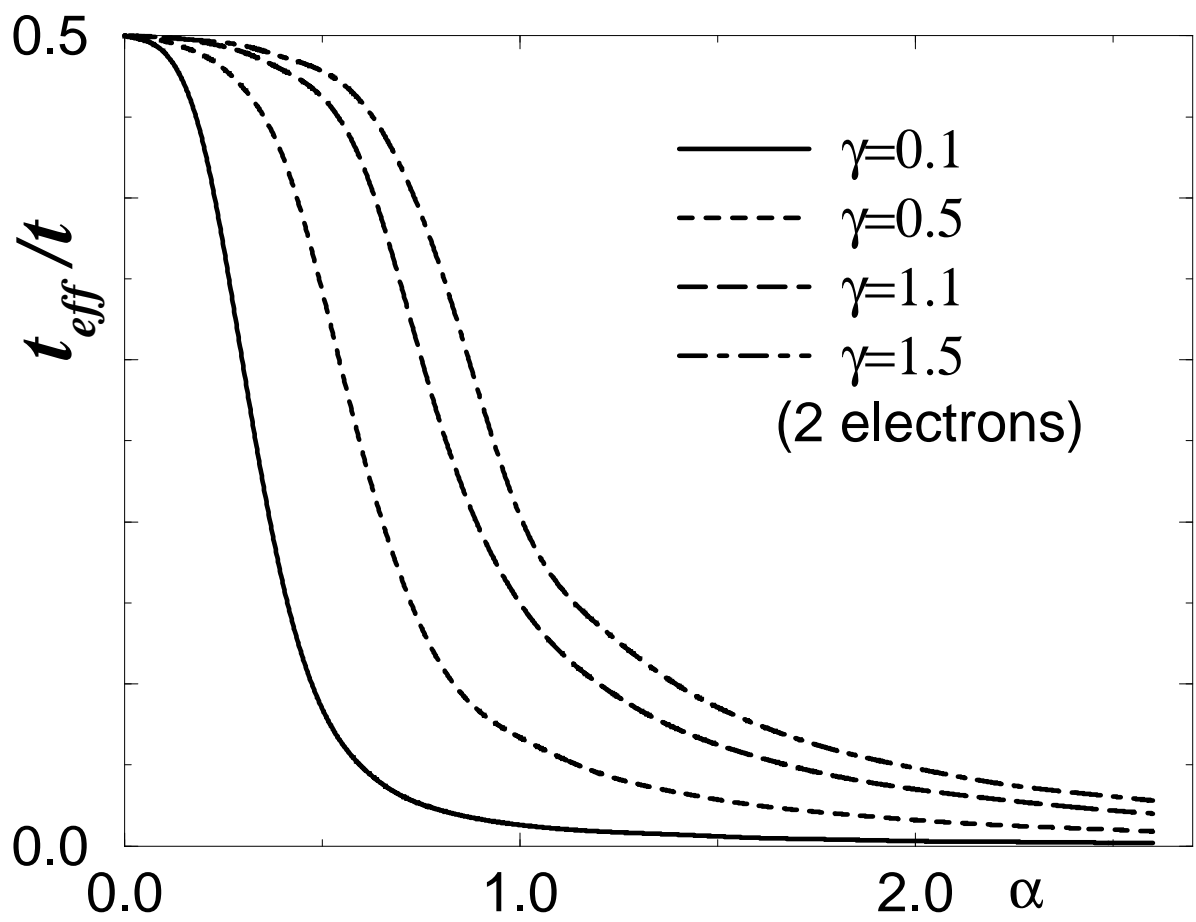
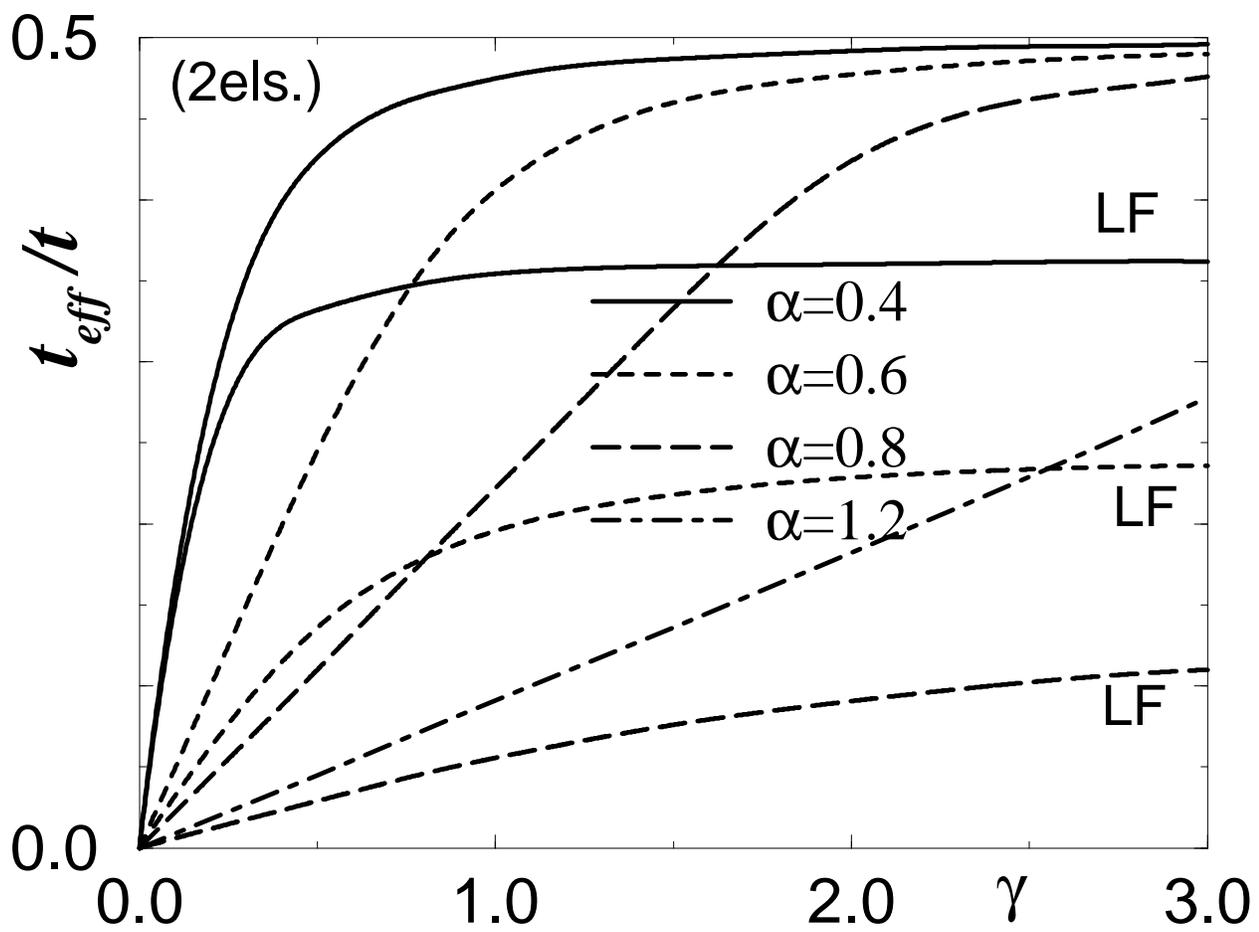
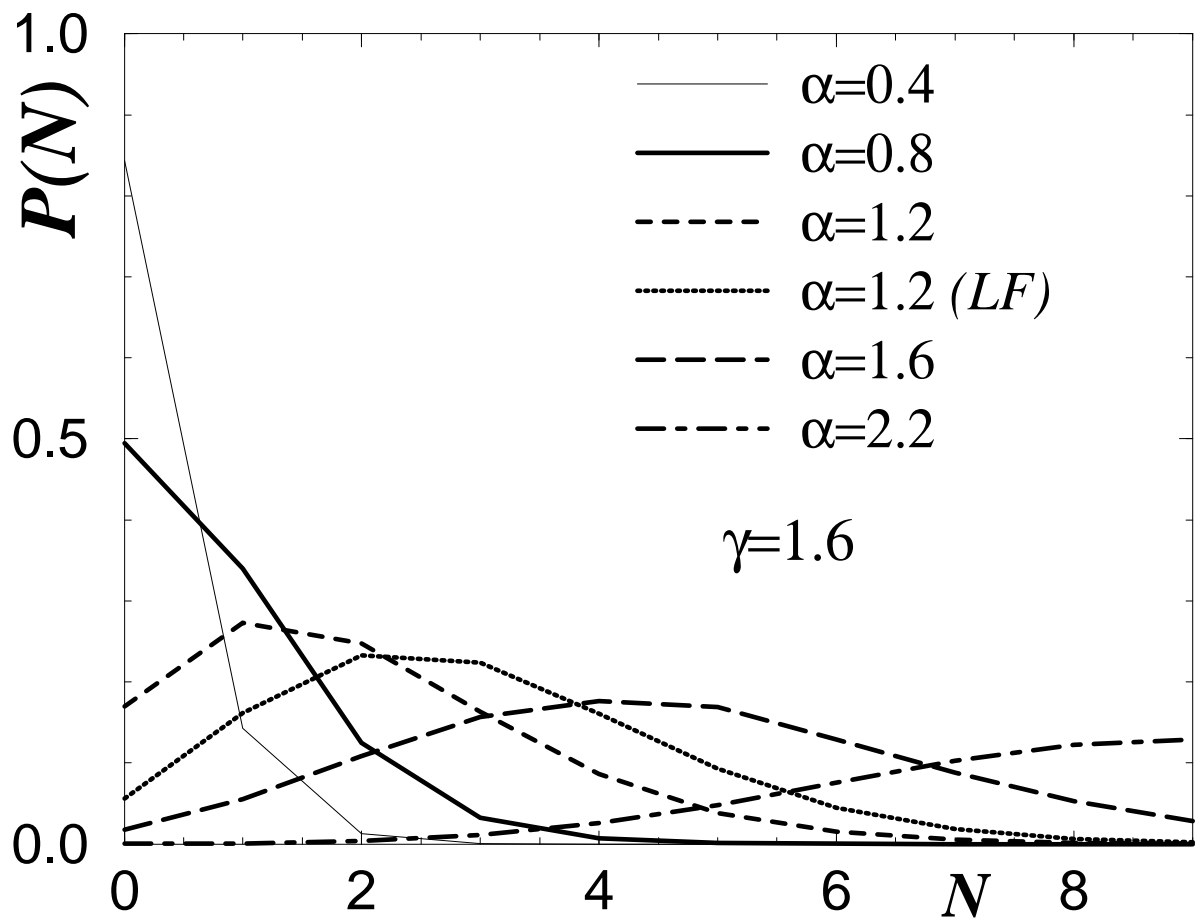
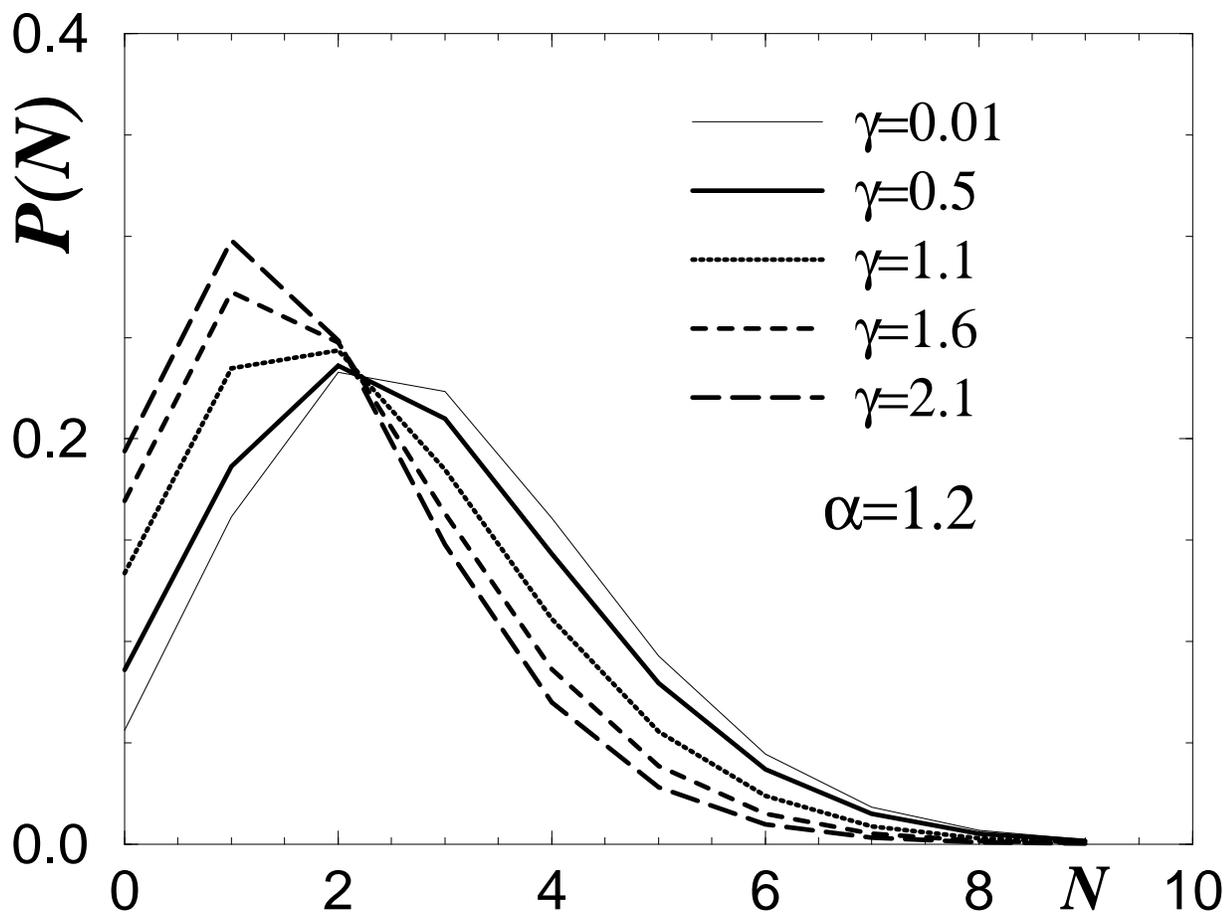
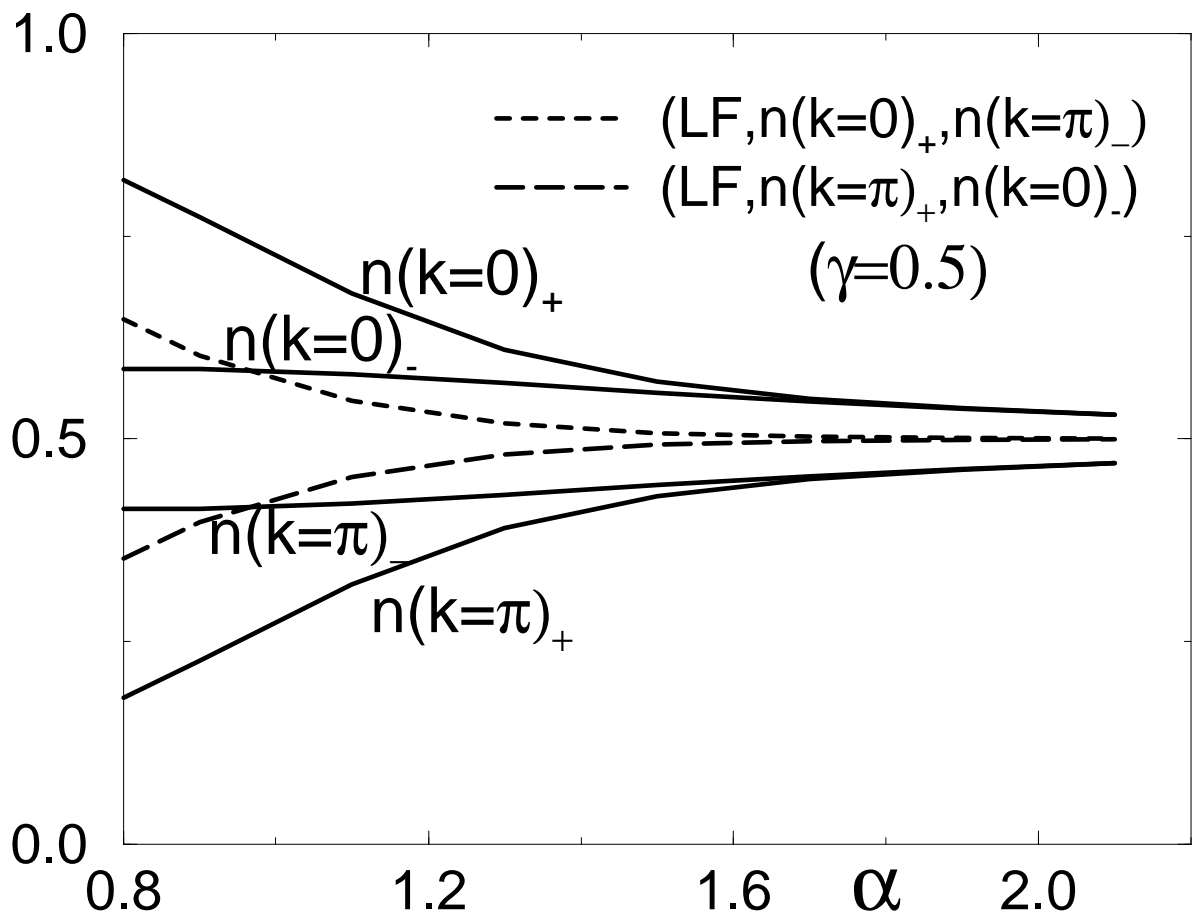
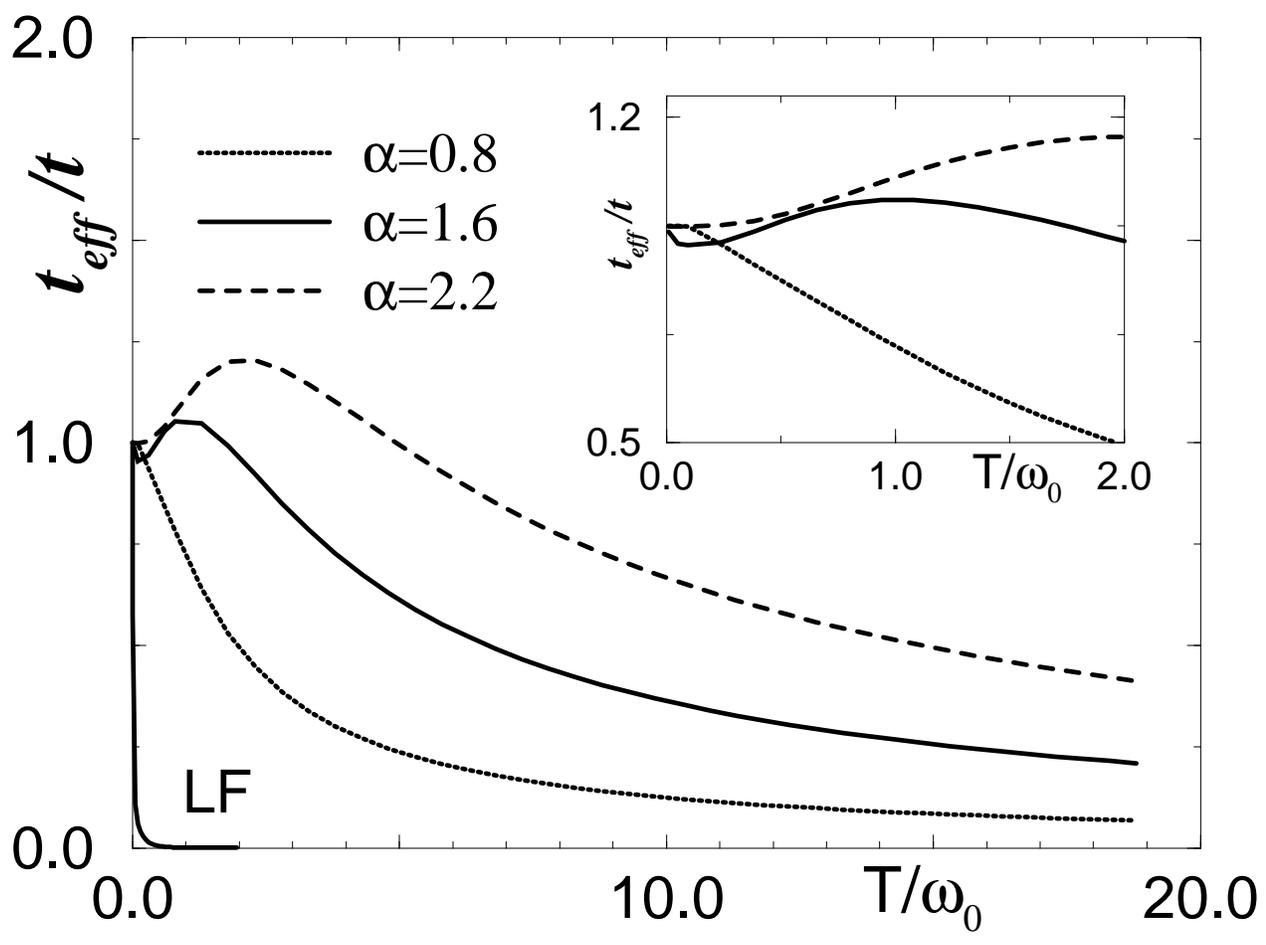
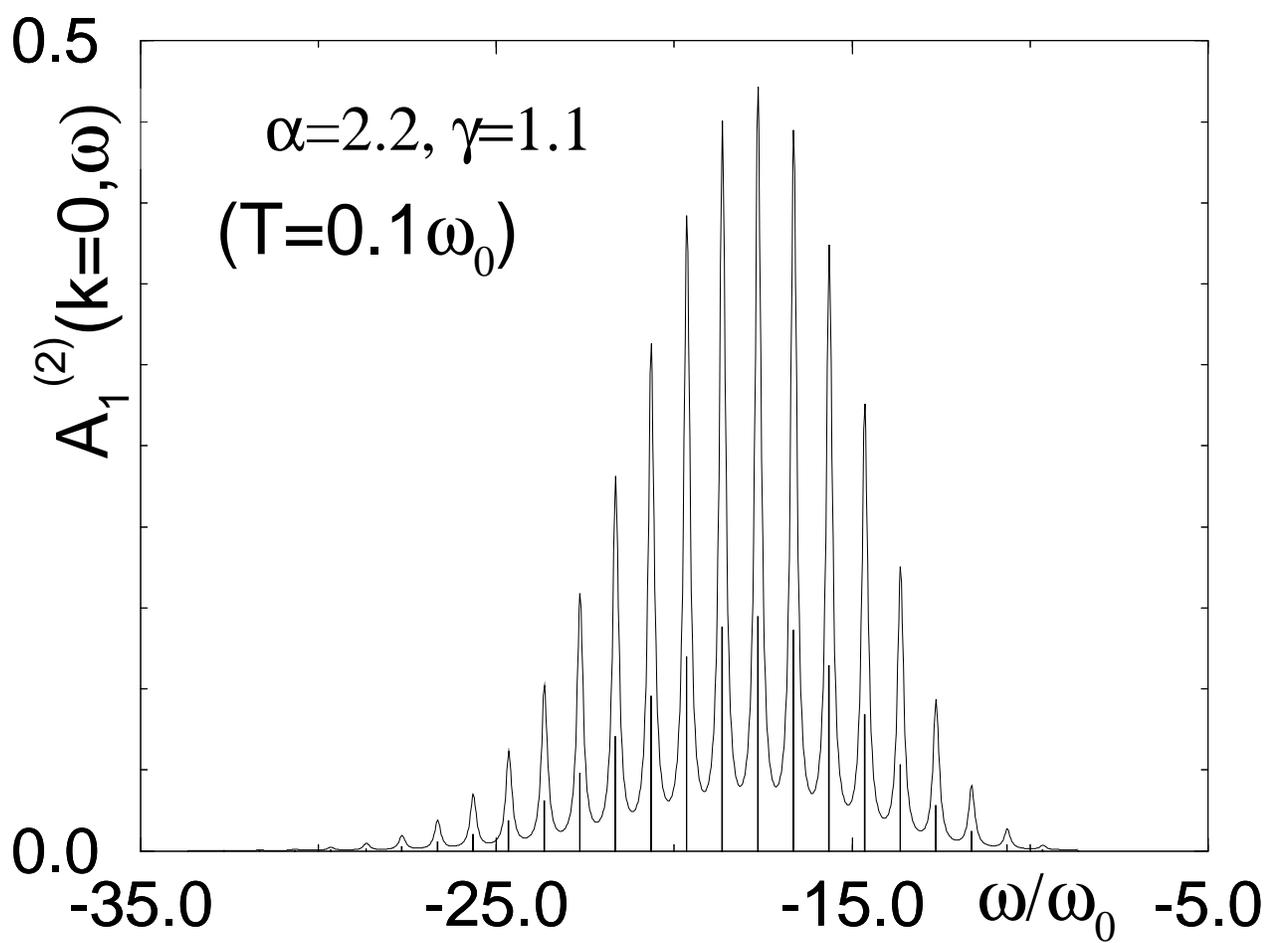
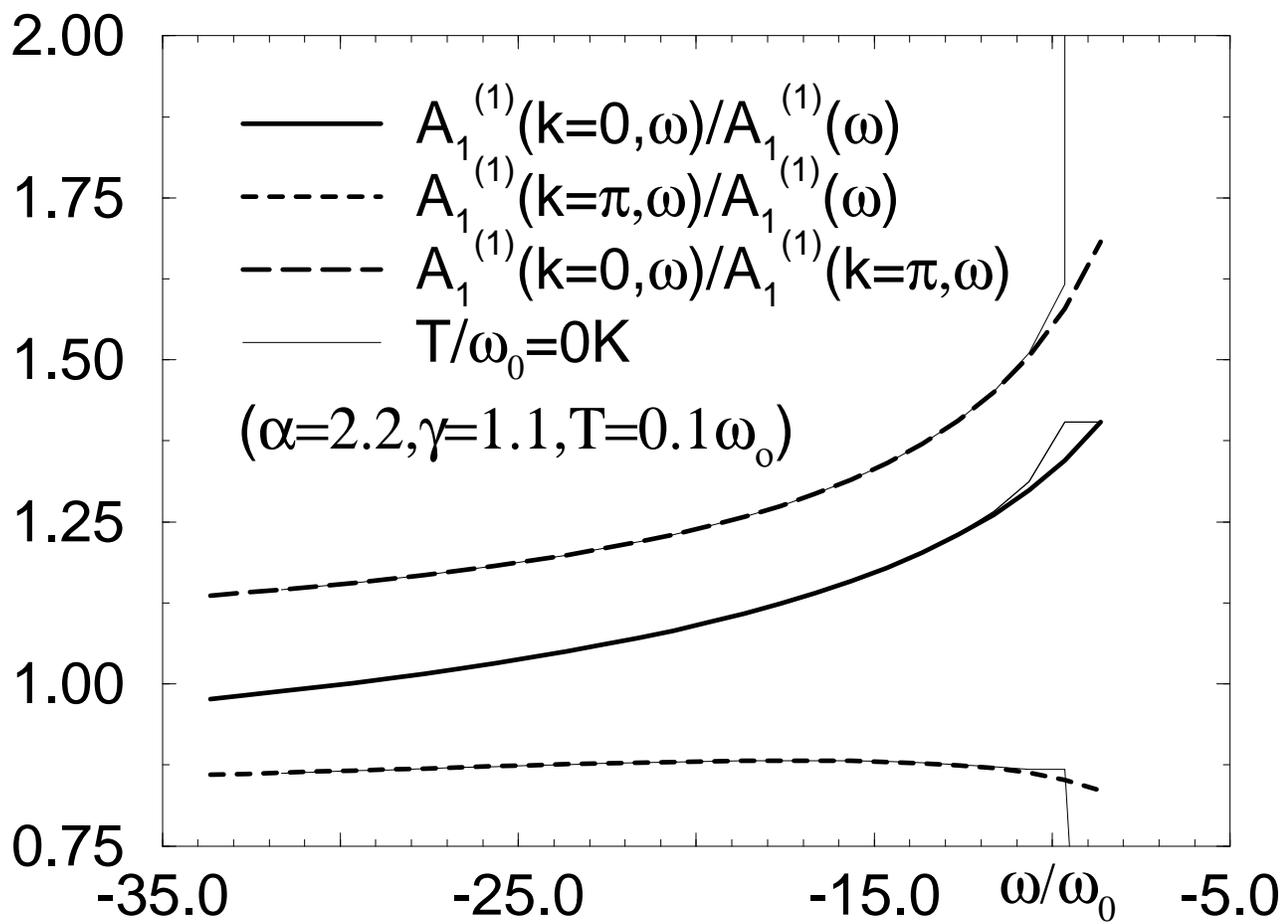
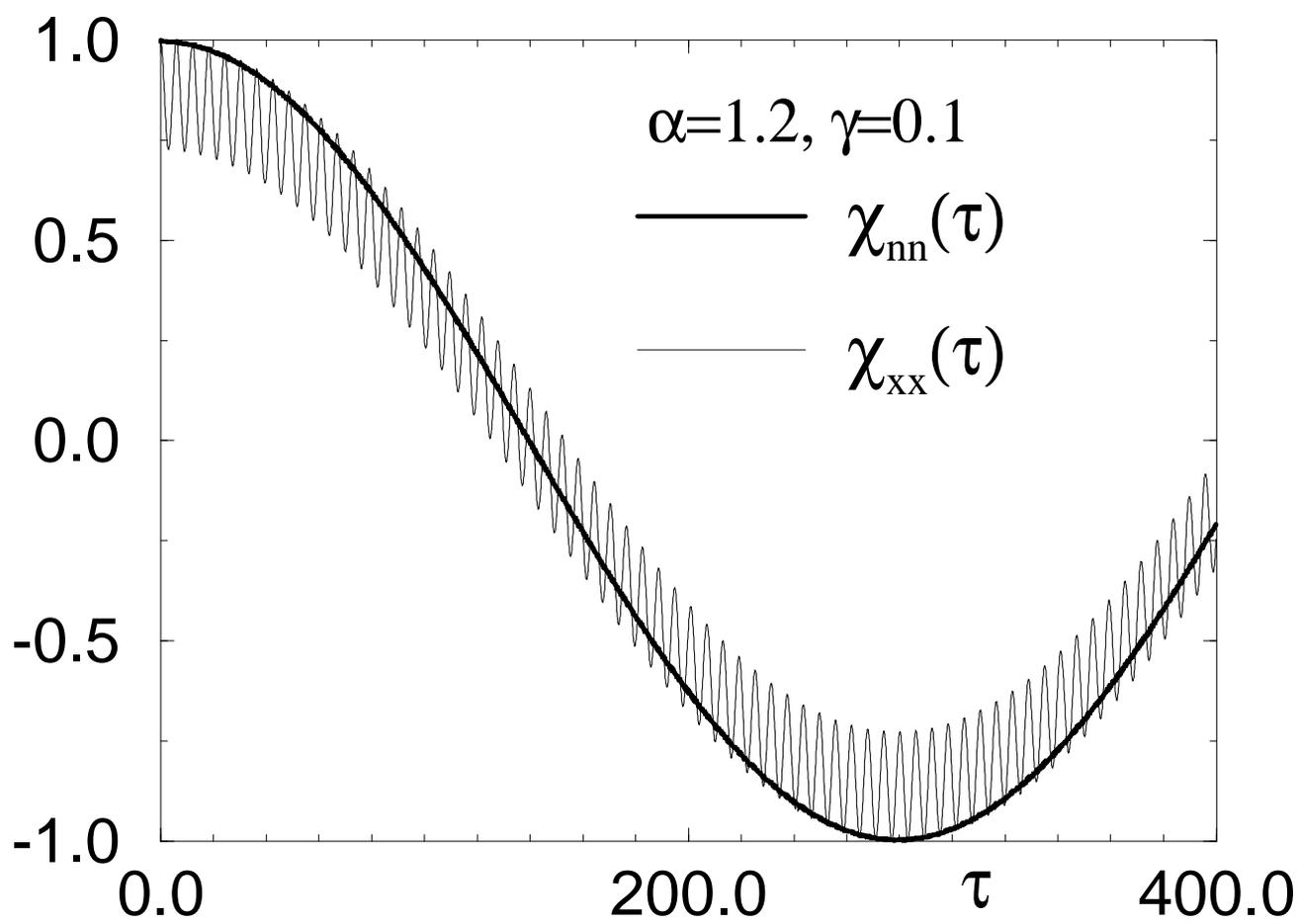
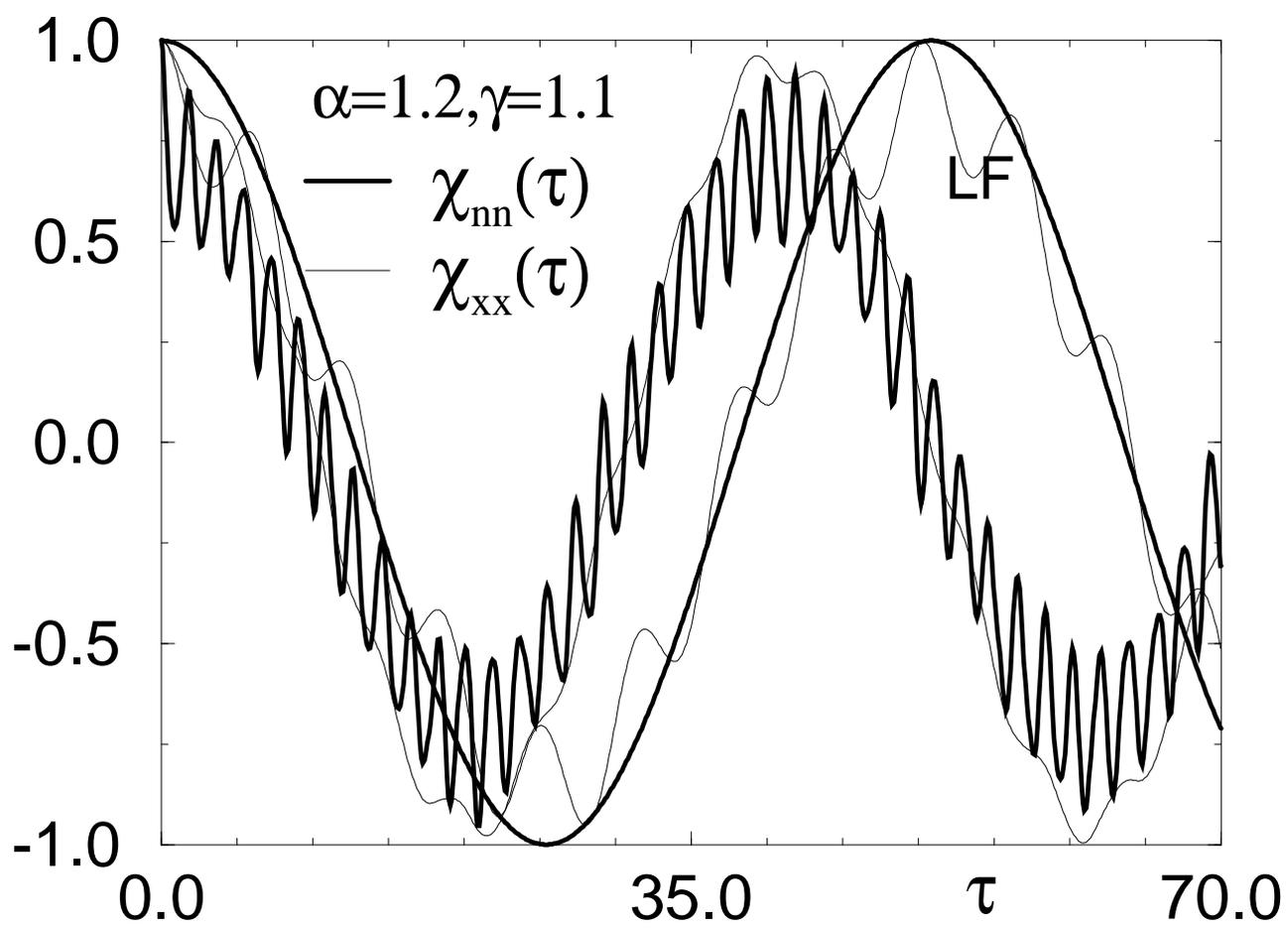
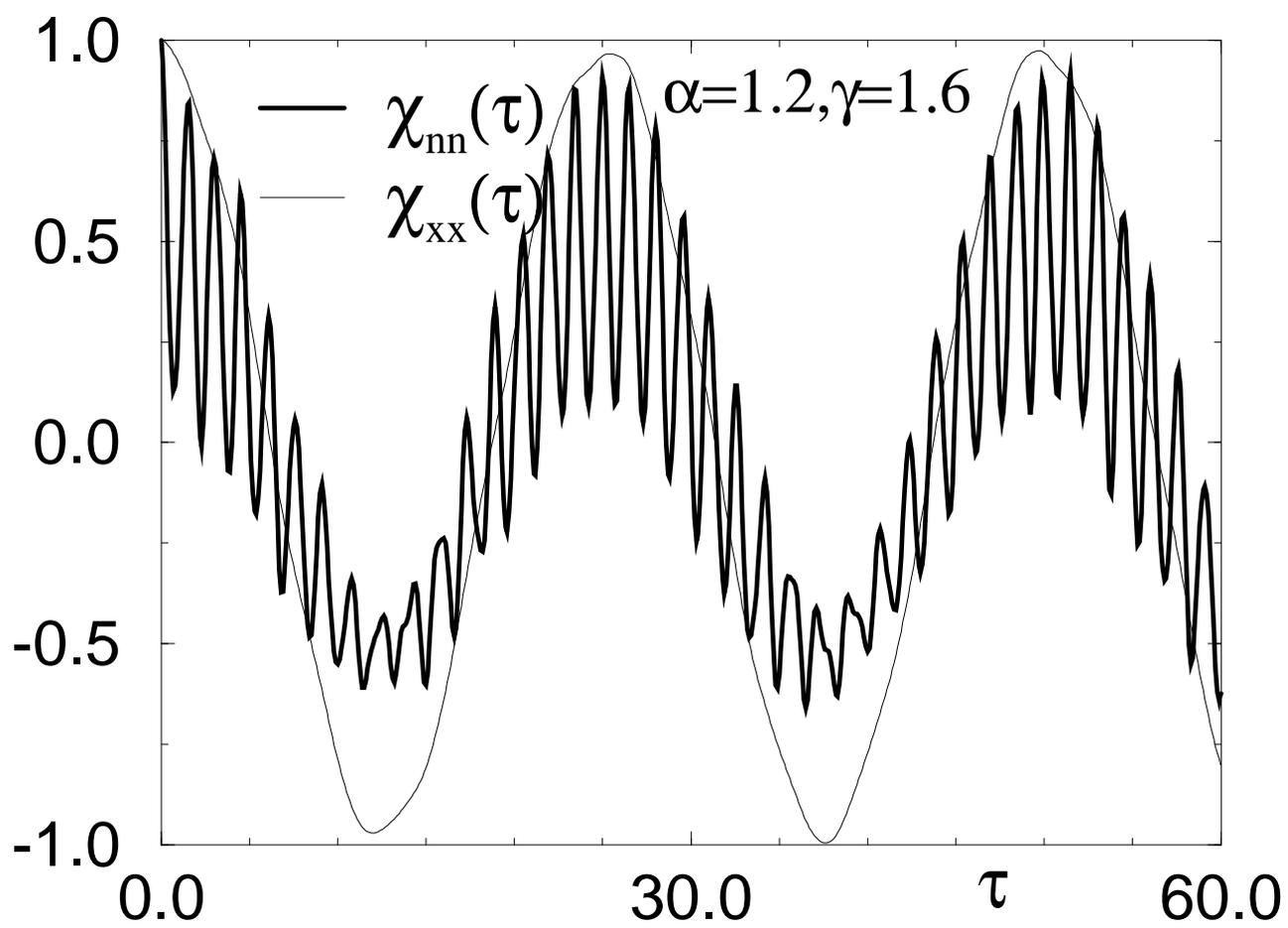
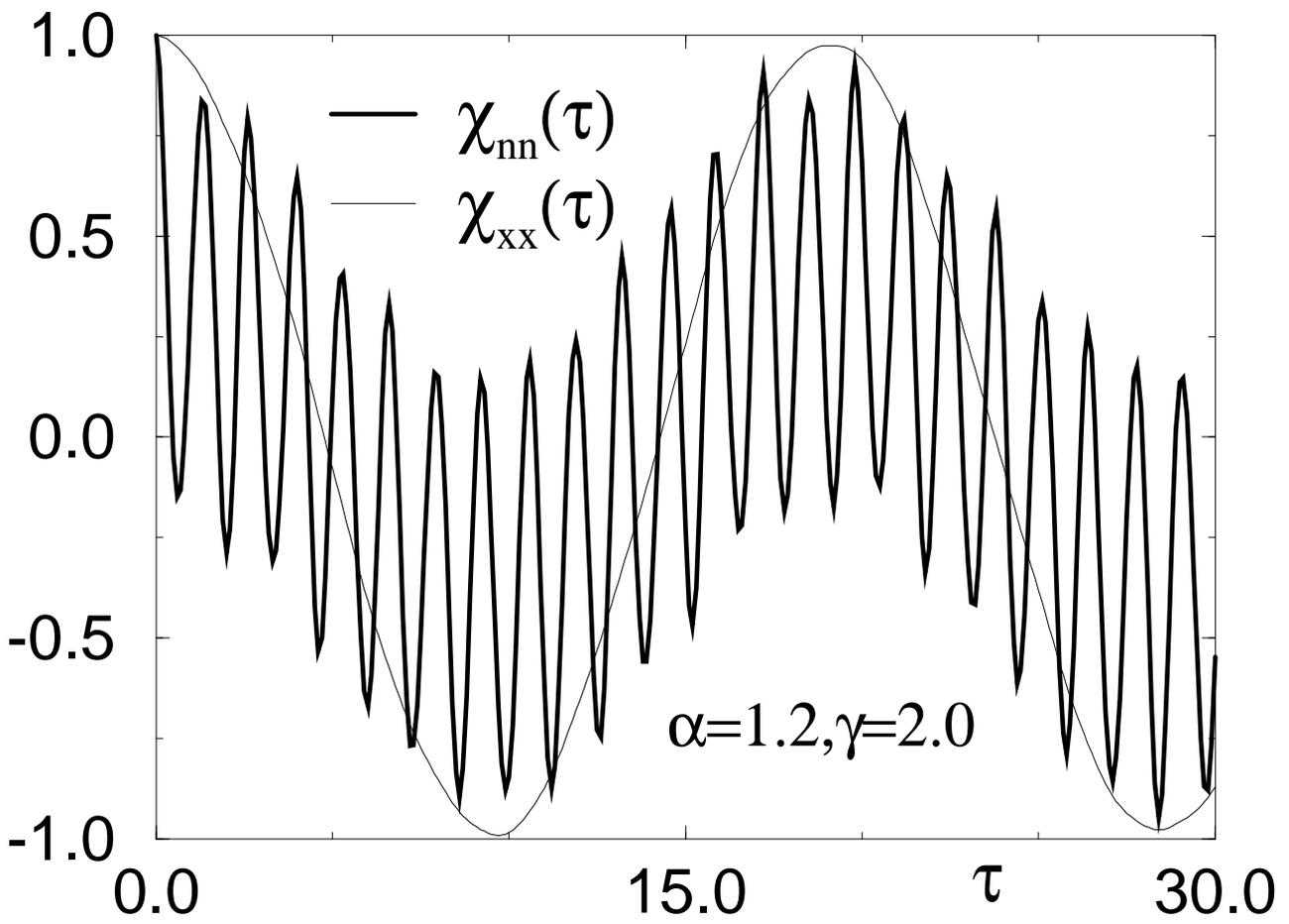
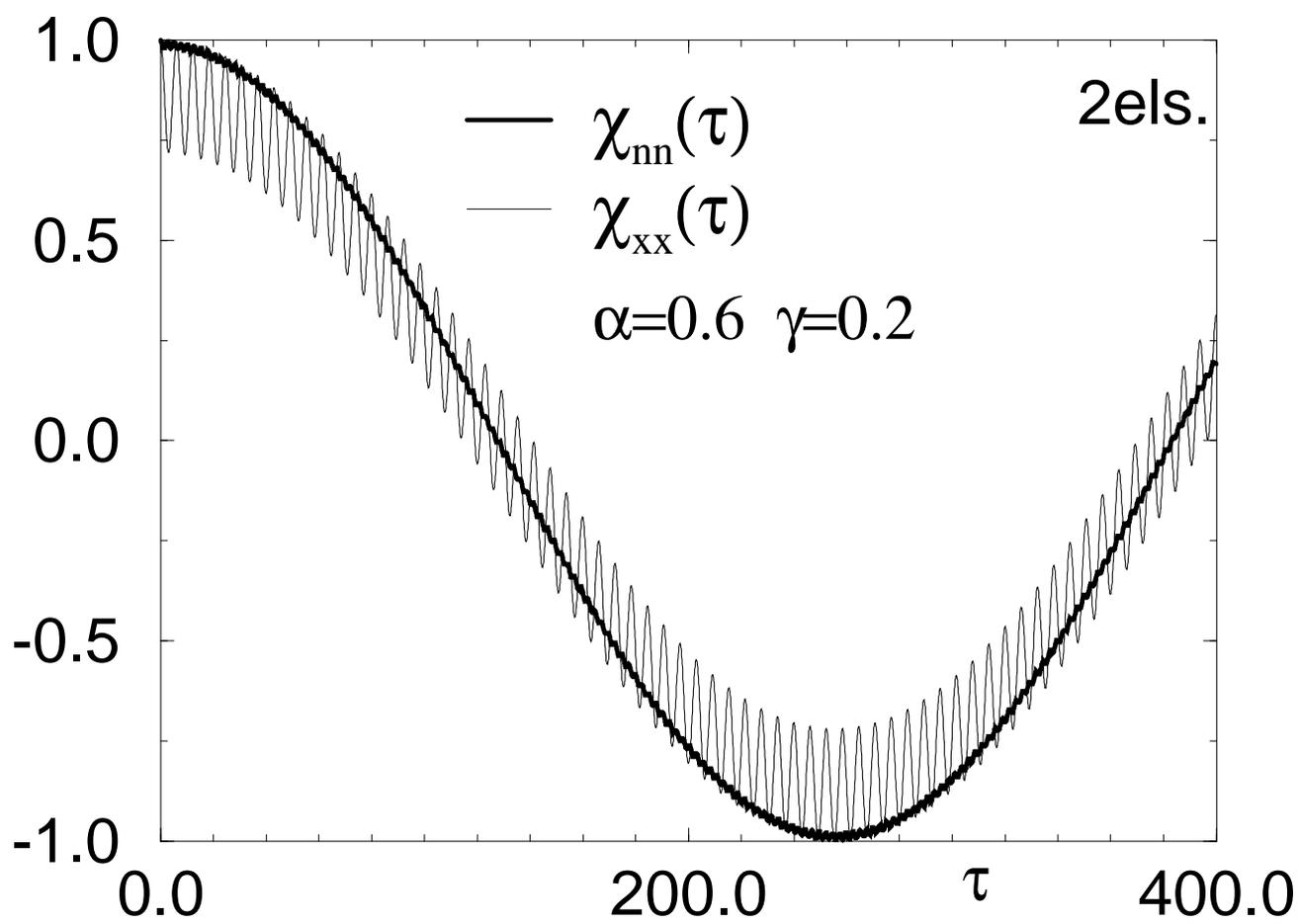
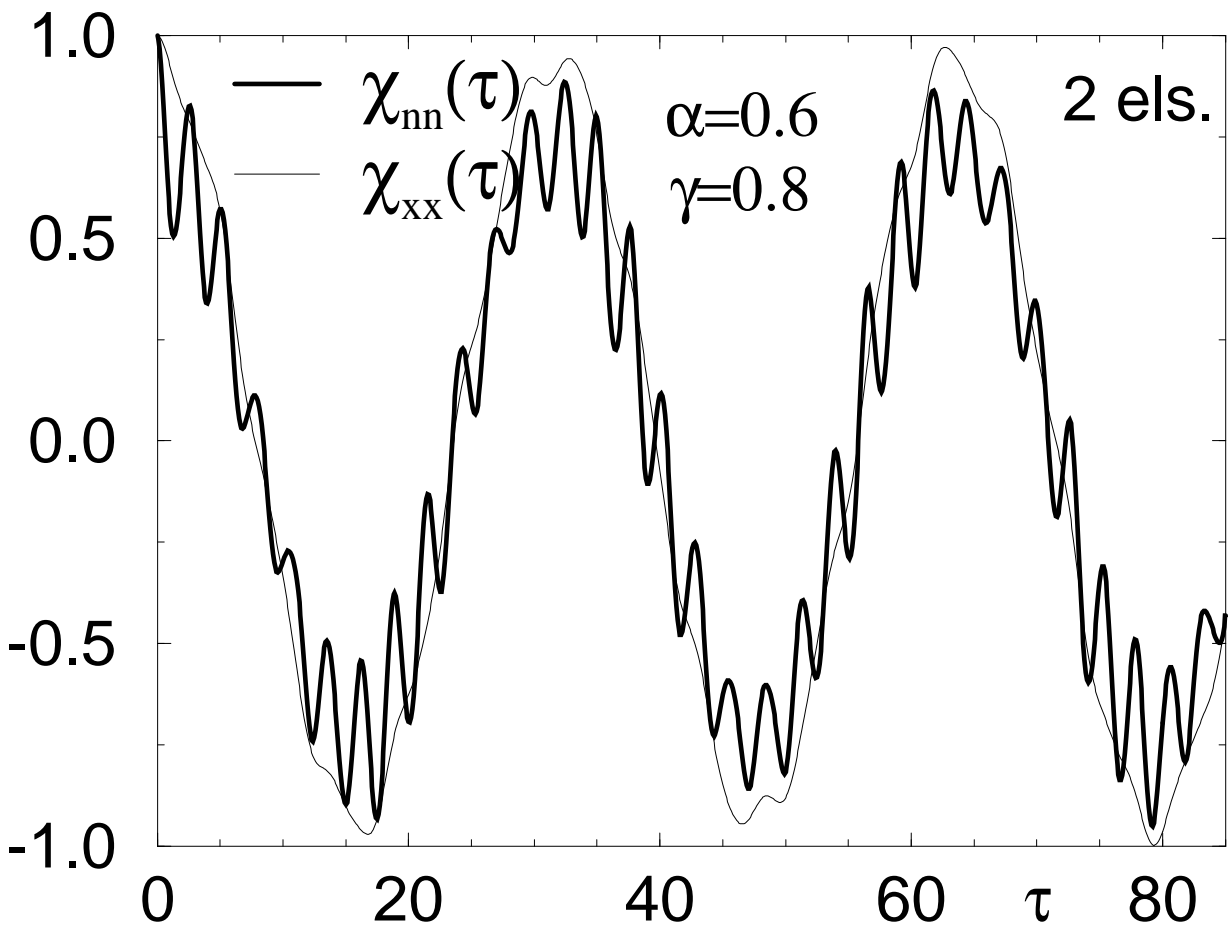
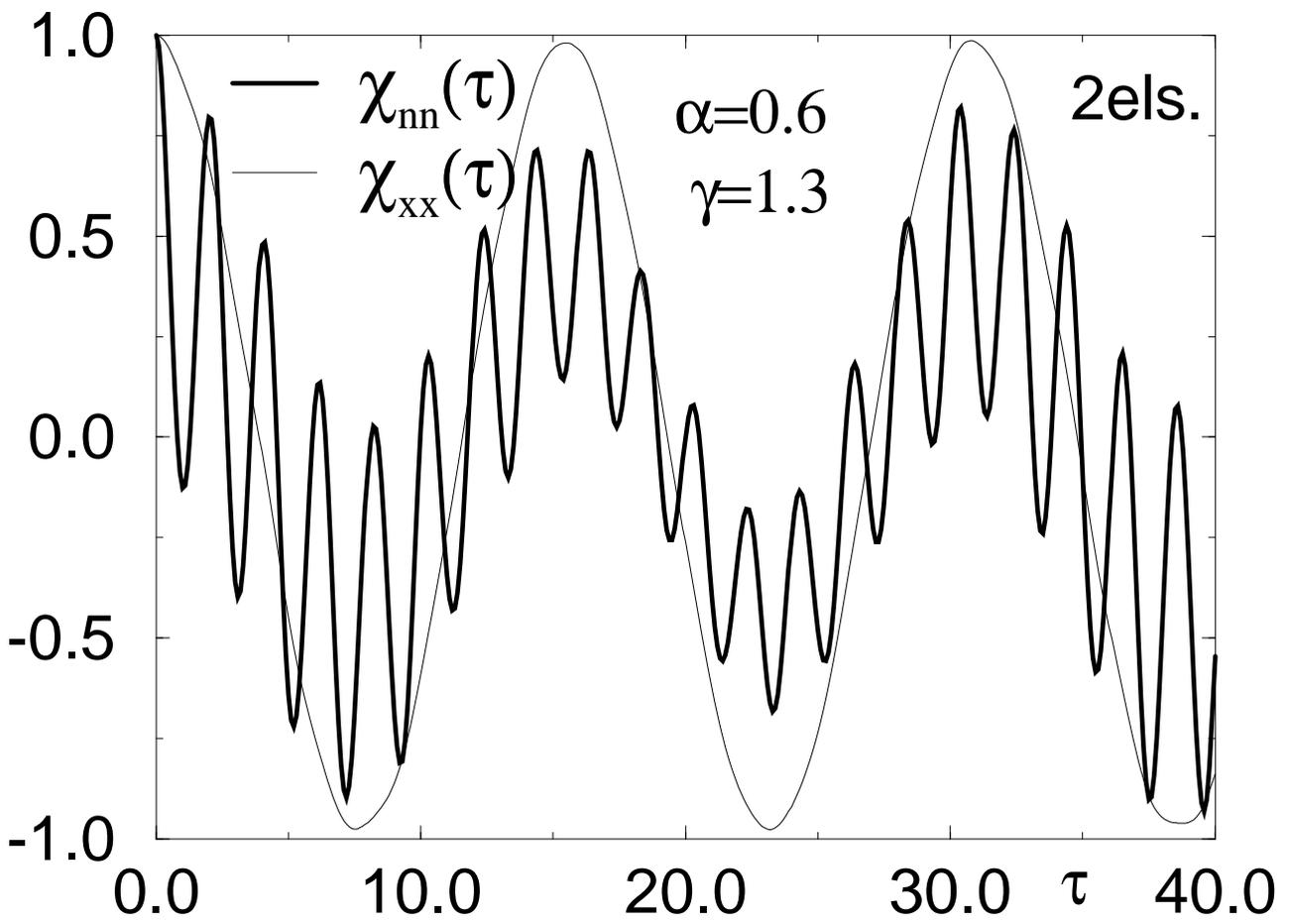
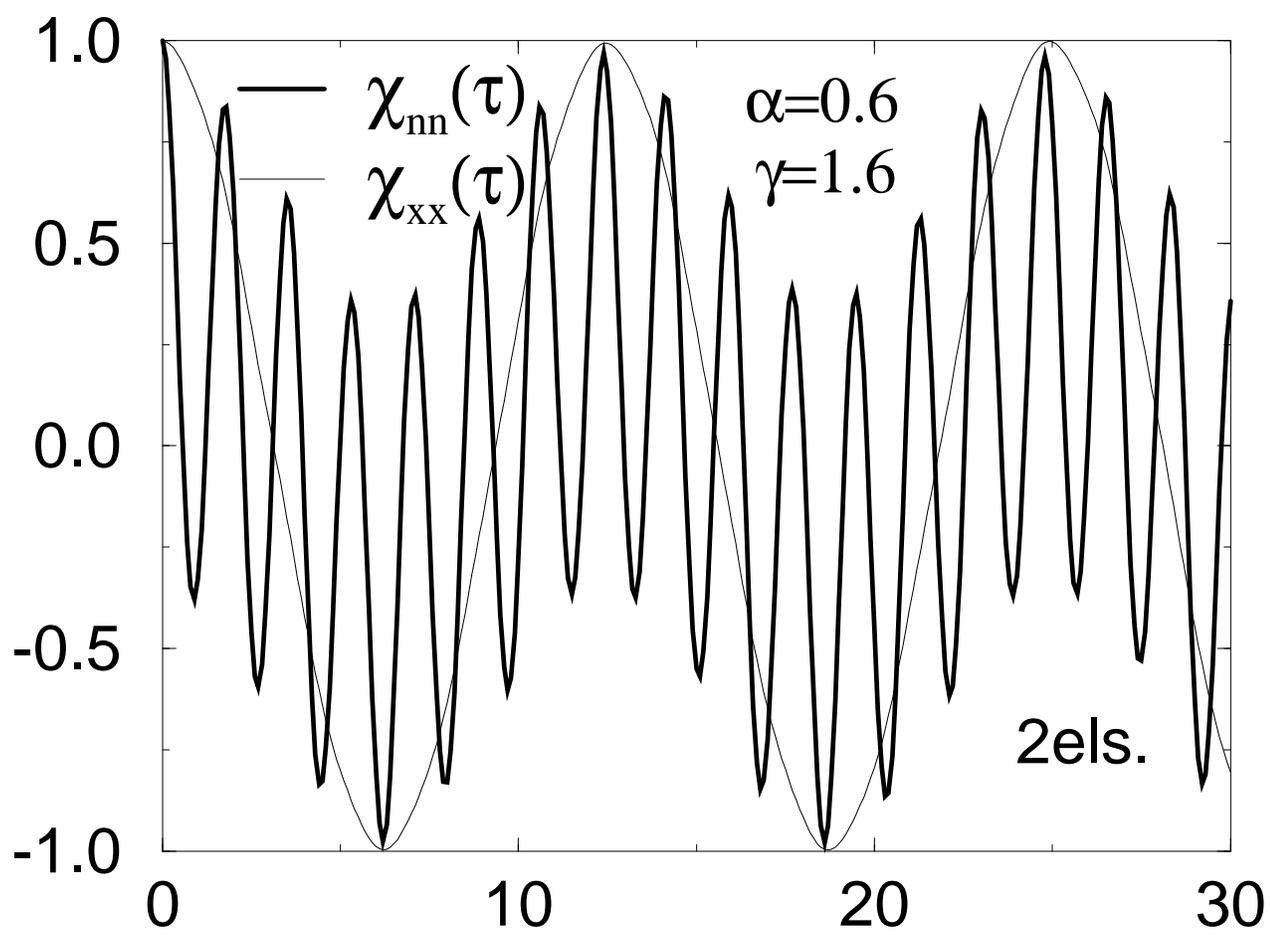
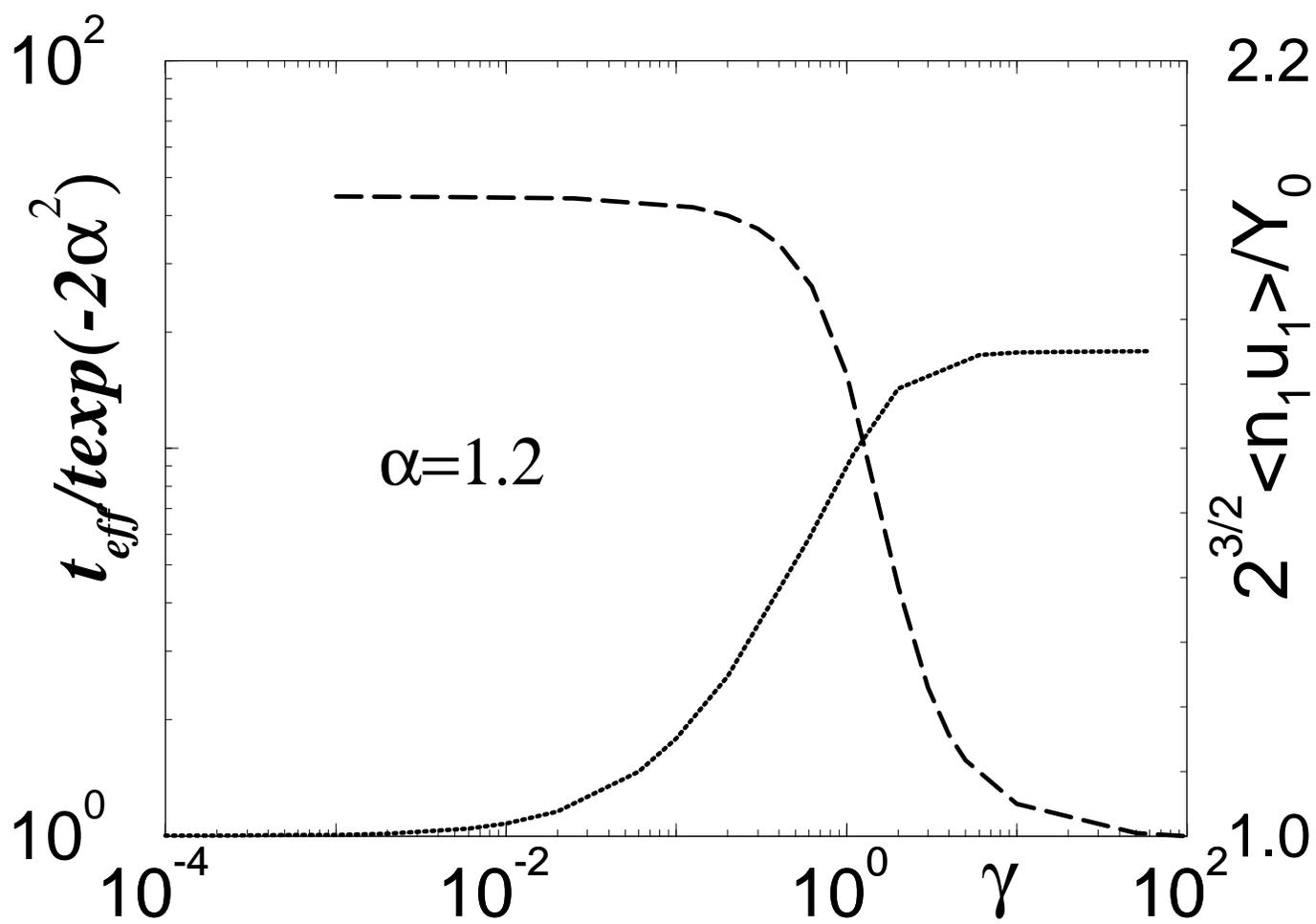
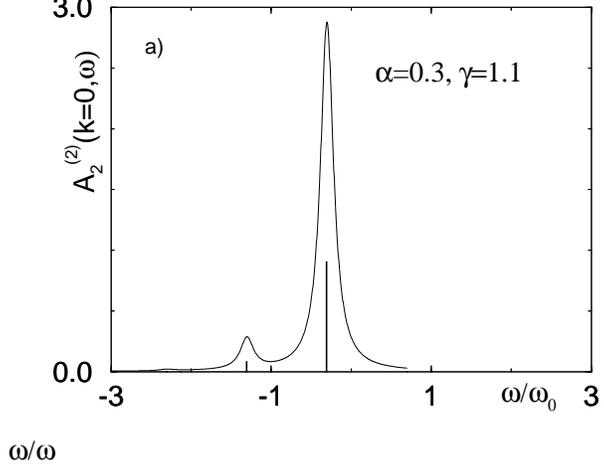
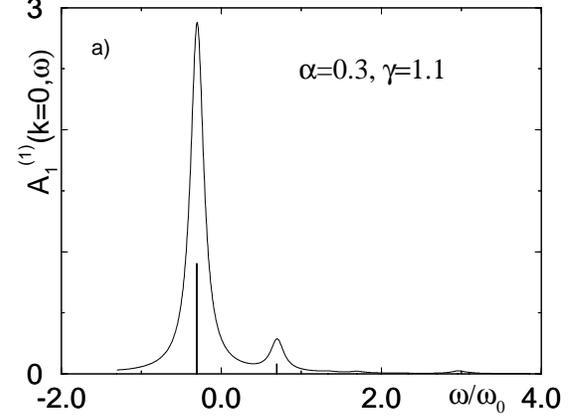
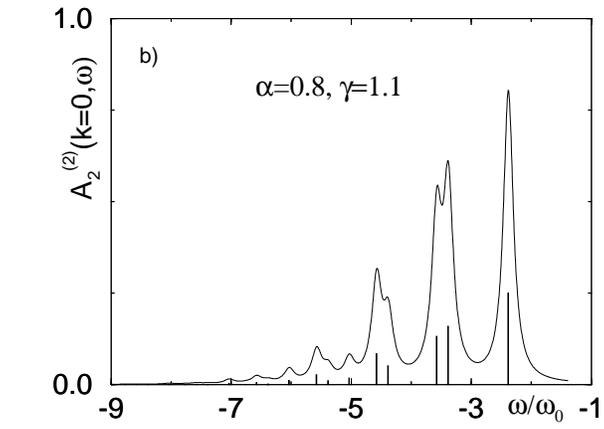
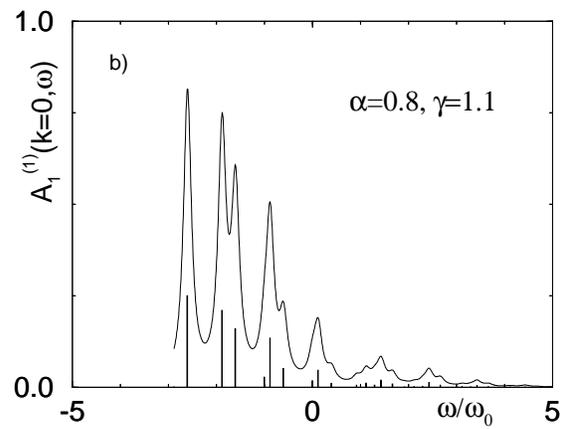
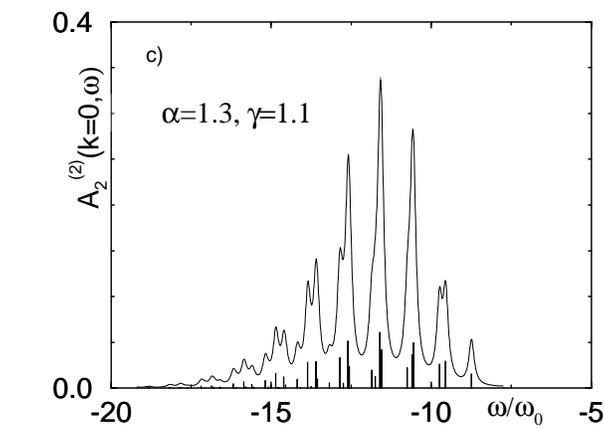
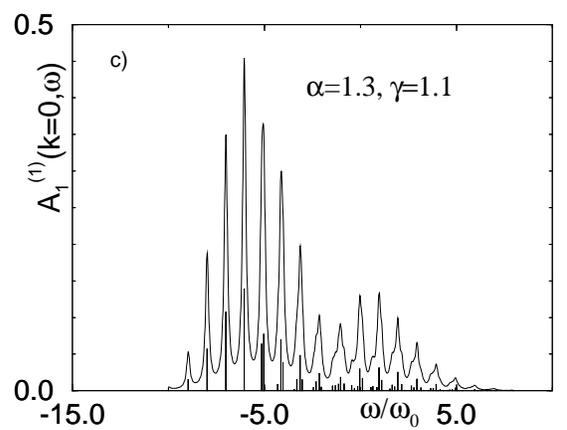
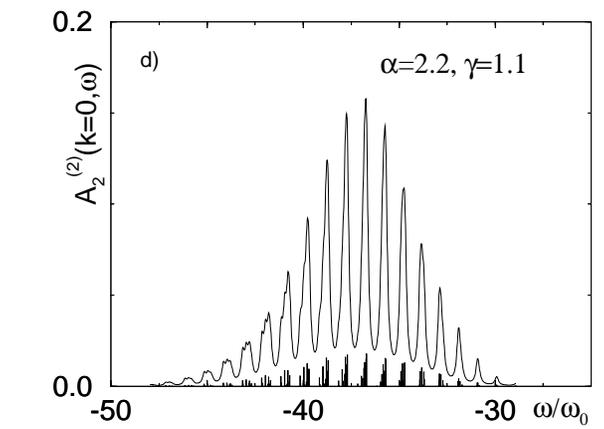
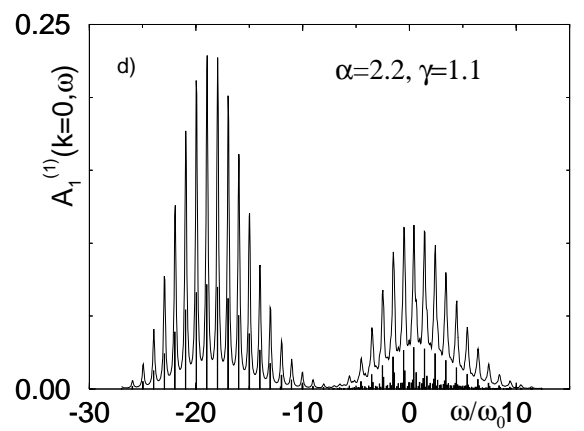
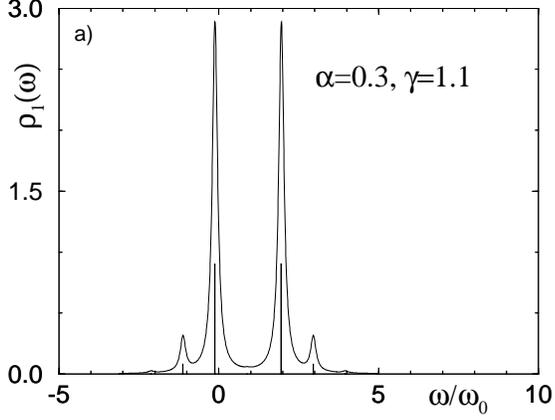
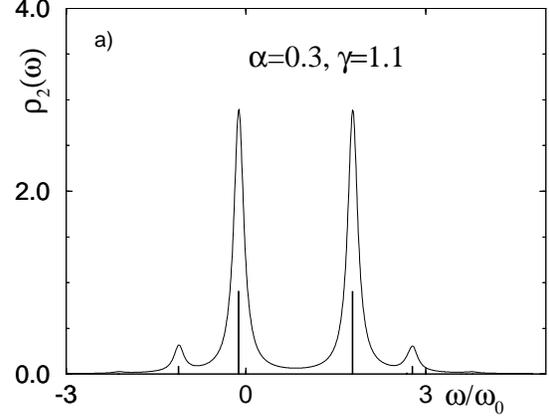
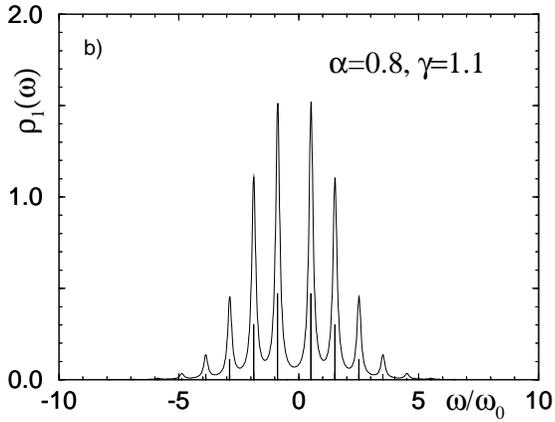
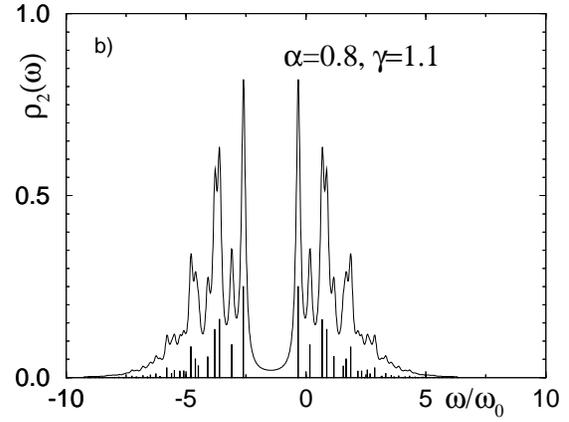
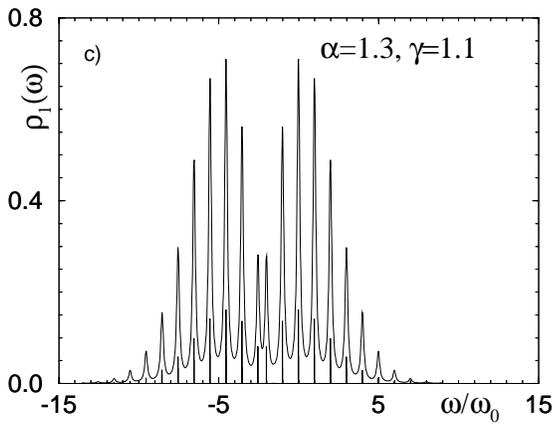
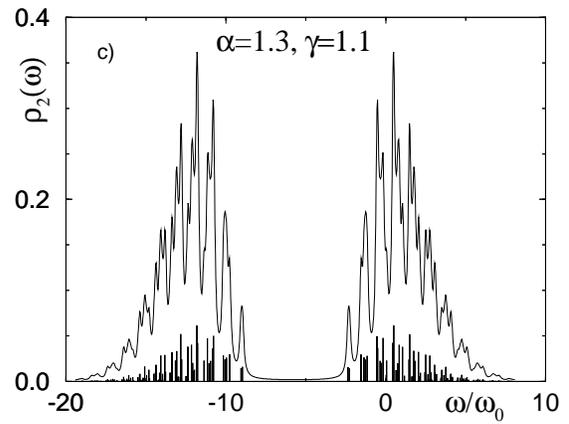
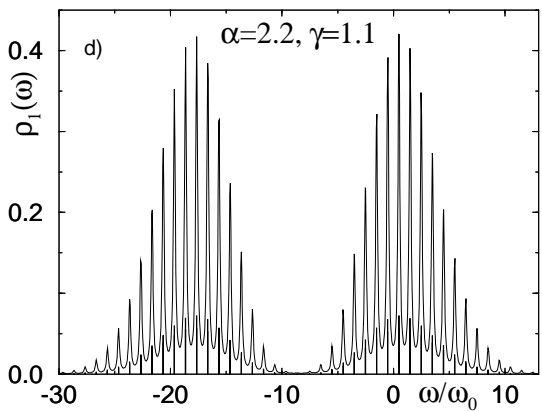
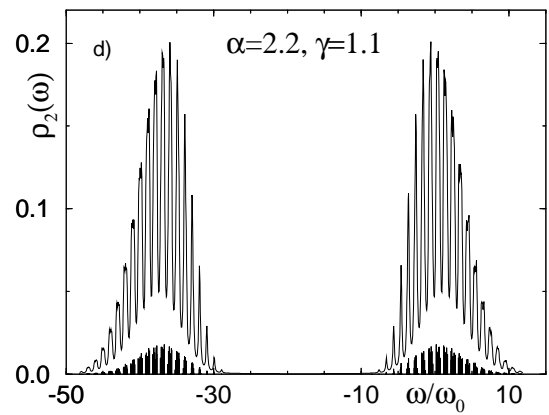
\end{document}